\documentclass[journal]{IEEEtran}
\usepackage{cite}
\usepackage{amsmath,amssymb,amsfonts}
\usepackage{algorithm,algorithmic}
\usepackage{graphicx}
\usepackage{textcomp}
\usepackage{xcolor}
\usepackage{multirow}
\usepackage[hidelinks]{hyperref}
\usepackage{array}
\usepackage{dblfloatfix}
\def\BibTeX{{\rm B\kern-.05em{\sc i\kern-.025em b}\kern-.08em
    T\kern-.1667em\lower.7ex\hbox{E}\kern-.125emX}}

\usepackage{xargs}
\usepackage[colorinlistoftodos,prependcaption,textsize=tiny]{todonotes}

\usepackage{csquotes}
\newcommand{\q}{\textquote}
\newcommand{\code}{\texttt}

\newcommand{\cost}[1]{\ensuremath{C_{\textrm{{#1}}}}}
\newcommand{\delay}[1]{\ensuremath{D_{\textrm{{#1}}}}}

\usepackage{enumitem}

\usepackage{orcidlink}

\usepackage{arydshln} 

\usepackage{svg}
\svgpath{{./img/}}

\begin{document}

\title{A Paradigm for Generalized\\Multi-Level Priority Encoders
}

\author{
Maxwell Phillips \orcidlink{0009-0003-5719-5523}, %
Firas Hassan \orcidlink{0009-0002-0387-7599}, and %
Ahmed Ammar \orcidlink{0000-0001-5907-7043} %
\thanks{Maxwell Phillips was with the Department of Electrical and Computer Engineering and Computer Science, Ohio Northern University, Ada, OH 45810, USA. He is now with the Department of Computer Sciences, University of Wisconsin--Madison, Madison, WI 53706 USA. Contact: mphillips@cs.wisc.edu.}%
\thanks{Firas Hassan and Ahmed Ammar are with the Department of Electrical and Computer Engineering and Computer Science, Ohio Northern University, Ada, OH 45810 USA. Contact: \{f-hassan, a-ammar\}@onu.edu.}%
}%

\markboth{}%
{
A Paradigm for Generalized Multi-Level Priority Encoders}

\maketitle

\begin{abstract}
Priority encoders are typically considered expensive hardware components in terms of complexity, especially at high bit precisions or input lengths (e.g., above 512 bits). However, if the complexity can be reduced, priority encoders can feasibly accelerate a variety of key applications, such as high-precision integer arithmetic and content-addressable memory. We propose a new paradigm for constructing priority encoders by generalizing the previously proposed two-level priority encoder structure. We extend this concept to three and four levels using two techniques---cascading and composition---and discuss further generalization. We then analyze the complexity and delay of new and existing priority encoder designs as a function of input length, for both FPGA and ASIC implementation technologies. In particular, we compare the multi-level structure to the traditional single-level priority encoder structure, a tree-based design, a recursive design, and the two-level structure. We find that the two-level architecture provides balanced performance -- reducing complexity by around half, but at the cost of a corresponding increase in delay. Additional levels have diminishing returns, highlighting a tradeoff between complexity and delay. Meanwhile, the tree and recursive designs are generally faster, but are more complex than the two-level and multi-level structures. We explore several characteristics and patterns of the designs across a wide range of input lengths. We then provide recommendations on which architecture to use for a given input length and implementation technology, based on which design factors---such as complexity or delay---are most important to the hardware designer. With this overview and analysis of various priority encoder architectures, we provide a priority encoder toolkit to assist hardware designers in creating the most optimal design.
\end{abstract}

\begin{IEEEkeywords}
high precision, multi-level, priority encoder, CLZ, CTZ, FFS, hardware acceleration
\end{IEEEkeywords}

\section{Introduction}\label{sec:intro}

\IEEEPARstart{A}{}priority encoder (PE) is a digital circuit that essentially identifies the highest-priority active input among several input signals, of which more than one may be active simultaneously. This basic functionality can be described in multiple equivalent ways. In the context of binary data, a PE can be said to find the most significant high bit (MSHB). From a mathematical perspective, for an input $x$, a PE finds $\lfloor\log_2 x\rfloor$. Furthermore, a PE with $n$ inputs has $\lfloor\log_2 n\rfloor$ outputs. More programmatically, a PE can also be used to implement the \textit{count leading zeroes} (CLZ) operation, also known as \textit{find first set} (FFS) or \textit{number of leading zeroes} (NLZ). With minor modifications, a PE can also replicate the related \textit{count trailing zeroes} (CTZ) or \textit{number of trailing zeroes} (NTZ) operations. 

These capabilities make priority encoders useful in a variety of applications, such as integer arithmetic \cite{2ldr, 2lmr}, multiplication overflow detection \cite{warren}, approximating $\log_2 x$ from $\lfloor \log_2 x\rfloor$ \cite{hassan, armsdg}, approximating $\sqrt{x}$ \cite{armsdg}, normalizing integers or converting them to floating-point numbers \cite{armsdg, warren}, binary and ternary content-addressable memory (particularly for networking and packet classification) \cite{bcam, tcam1, tcam2}, incrementers and decrementers \cite{tcam3, tree, xtn1, xtn3}, information retrieval \cite{xtn1}, and data compression \cite{int_compress}. 
Several of these specific operations play key roles in broader areas such as encryption and hardware acceleration of cryptographic algorithms. In particular, the discrete logarithm problem is essential to elliptic curve cryptography (ECC) \cite{ecc}, and large integer multiplication and division are fundamental operations in RSA encryption \cite{2lmr, 2ldr, eval-rafferty, rsa_sutter, ecc}.

The primary obstacle to widespread adoption of priority encoders is their high complexity and poor scalability. PEs have traditionally been considered expensive components that do not scale well with increasing input length (or bit precision) $n$. As such, they can be impractical for \q{high-precision} applications where $n>512$, as the cost and complexity at this bit precision quickly become prohibitive \cite{2ldr, 2lmr, tcam2}. Motivated by these factors, our contributions are as follows:

\begin{enumerate}[label=\textbf{\arabic*.}]
    \item We introduce two techniques, \textit{cascading} and \textit{composition}, by which we extend the two-level priority encoder concept to multiple levels.
    \item We discuss several methods of constructing priority encoders to reduce hardware complexity (and by extension, power and cost).
    \item We perform a detailed analysis of and comparison between these methods, focusing on their relative complexity and delay.
\end{enumerate}

To the best of our knowledge, the existing literature does not present an analysis of this depth across different priority encoder structures.

\begin{figure*}[t!]
    \centering
    \vspace{-1em}
    \includegraphics[width=1.5\columnwidth]{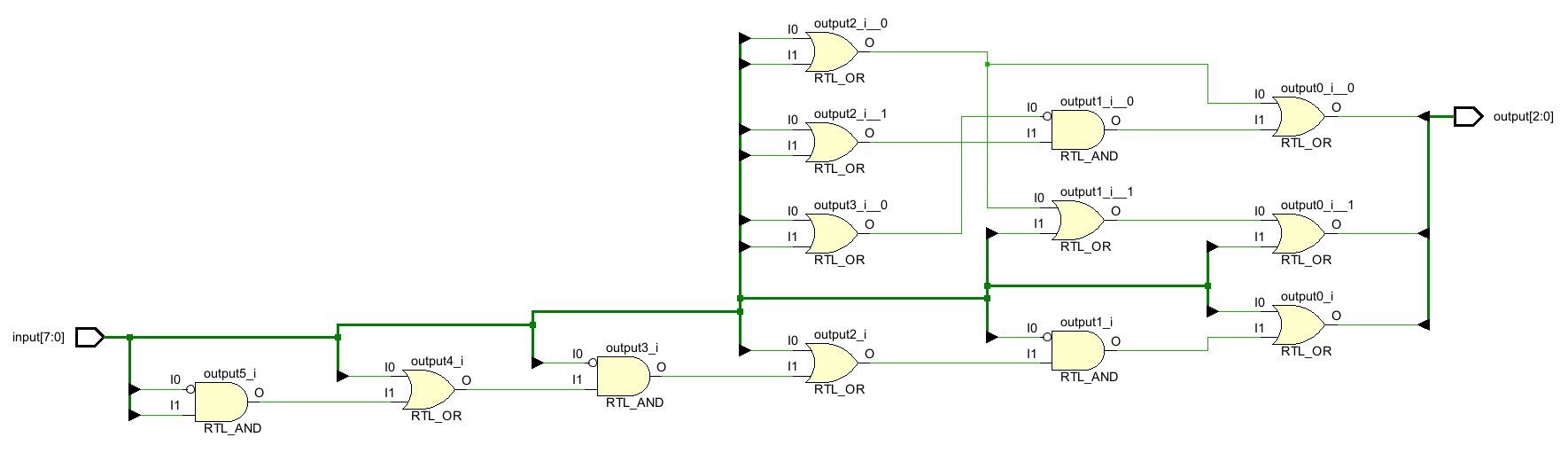}
    \vspace{-1em}
    \caption{A gate-based 8:3 PE.}
    \label{fig:slpe-8-gate}
    \vspace{-0.25em}
\end{figure*}

\begin{figure*}[t!]
    \centering
    \vspace{-1em}
    \includegraphics[width=2\columnwidth]{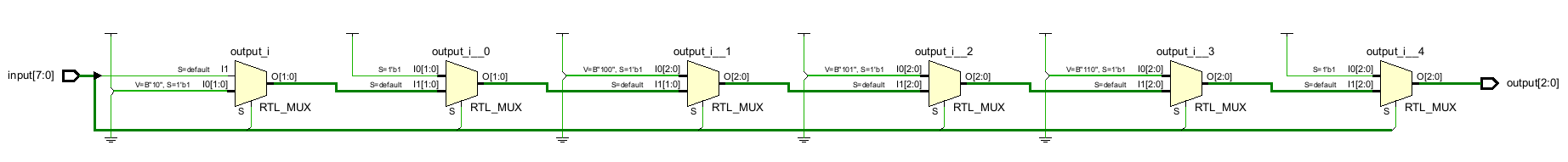}
    \vspace{-1em}
    \caption{A mux-based 8:3 PE.}
    \label{fig:slpe-8-mux}
    \vspace{-0.25em}
\end{figure*}

The remainder of this paper is structured as follows. Section~\ref{sec:bg} discusses the background and general characteristics of priority encoders as well as various methods of constructing them, distinct from our multi-level techniques. Section~\ref{sec:2lpe} introduces the two-level priority encoder and examines it in detail. Section~\ref{sec:mlpe} presents the main contribution of this paper: the extension and generalization of the two-level concept to multiple levels. Next, Section~\ref{sec:components} describes and analyzes the low-level components and structure of priority encoders, and Section~\ref{sec:methods} details our methodology for assembling these components into complete priority encoders. In Section~\ref{sec:results}, we summarize FPGA synthesis results and theoretical ASIC performance, then provide our recommendations on what architectures to choose based on design goals and constraints as well as implementation technology. Section~\ref{sec:conclusion} concludes the paper and outlines directions for future work.

\section{Background}\label{sec:bg}
Priority encoder designs may be characterized by the number of levels, type of recursion, and serial vs. parallel nature. In this section, we discuss existing designs, including the traditional single-level priority encoder and two different contemporary PE designs: \q{recursive} \cite{bcam} and \q{tree} \cite{tree}. This work focuses on parallel (or combinational) PEs, but we briefly comment on serial PEs near the end of this section. 

\subsection{Standard (Single-Level) Priority Encoders}\label{sec:slpe}
A standard or \textit{single-level} priority encoder (SLPE) is the most \q{basic} or \q{traditional} design, and the least scalable. At the hardware level, an SLPE can be implemented in two ways:
\begin{enumerate}[label=\textbf{\arabic*.}]
    \item by directly translating its Boolean logic expressions into digital logic gates (which we call \textit{gate-based}), as shown in Fig.~\ref{fig:slpe-8-gate};
    \item by constructing a chain of multiplexers (which we call \textit{mux-based}), as shown in Fig.~\ref{fig:slpe-8-mux}.
\end{enumerate}

In addition to the primary output, PEs often also provide a \textit{valid} signal, which provides a method for distinguishing between the states of \q{MSHB at bit 0, the least significant bit (LSB)} and an input of all zeroes, which is typically considered an invalid input. The valid output, usually a single bit generated by ORing all inputs, is \code{0} only when the input is all zeroes, and \code{1} at any other time (assuming active high outputs).

\subsection{Recursive Priority Encoders}\label{sec:recursive_pe}
A recursive structure%
\footnote{Of course, this is not the \textit{only} recursive structure for a PE; the tree and multi-level structures discussed later in this work are also recursive in nature. We use this notation to be consistent with \cite{bcam} and to differentiate it from \cite{tree}.} 
for priority encoders, introduced by \cite{bcam}, is shown in Fig.~\ref{fig:recursive} below.%
\footnote{Published in the original work \cite{bcam} and also separately via Wikimedia Commons under CC-BY-SA 4.0 \cite{bcam-fig}.}
In this design, the PE is split into $k$ sub-encoders of input length $n/k$. The \textit{valid} signal outputs from these feed a smaller PE of input length $k$. This PE is then used to select one of the outputs of the first layer of PEs via a multiplexer.

\begin{figure}[ht]
    \centering
    \vspace{-0.75em}
    \includesvg[inkscapelatex=false, width=\columnwidth]{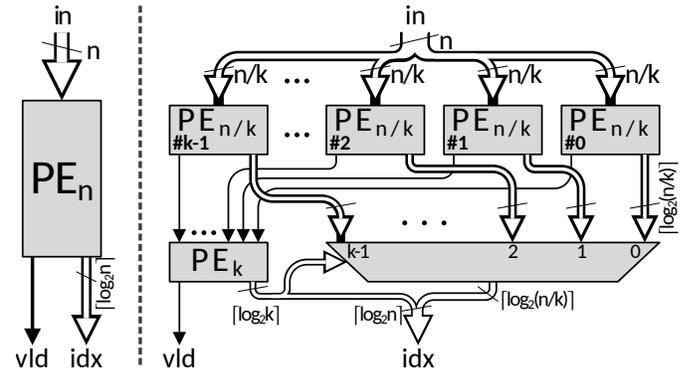}
    \vspace{-0.75em}
    \caption{Recursive PE Structure \cite{bcam, bcam-fig}.}
    \label{fig:recursive}
    \vspace{-0.25em}
\end{figure}

This structure reduces the complexity of the priority encoder by a constant factor. Notably, as we found in our replication of this design, it manages this \textit{without increasing the delay} compared to an SLPE.%
\footnote{See Section~\ref{sec:results} for more details.}
Furthermore, as described in \cite{bcam}, this structure is highly conducive to implementation on FPGAs when $k=4$, since a 4:1 mux can be implemented on a single 6-input lookup table (LUT6), common in many FPGAs.

\subsection{Tree Priority Encoders}\label{sec:tree_pe}
\begin{figure}[ht]
    \centering
    \includegraphics[width=0.95\columnwidth]{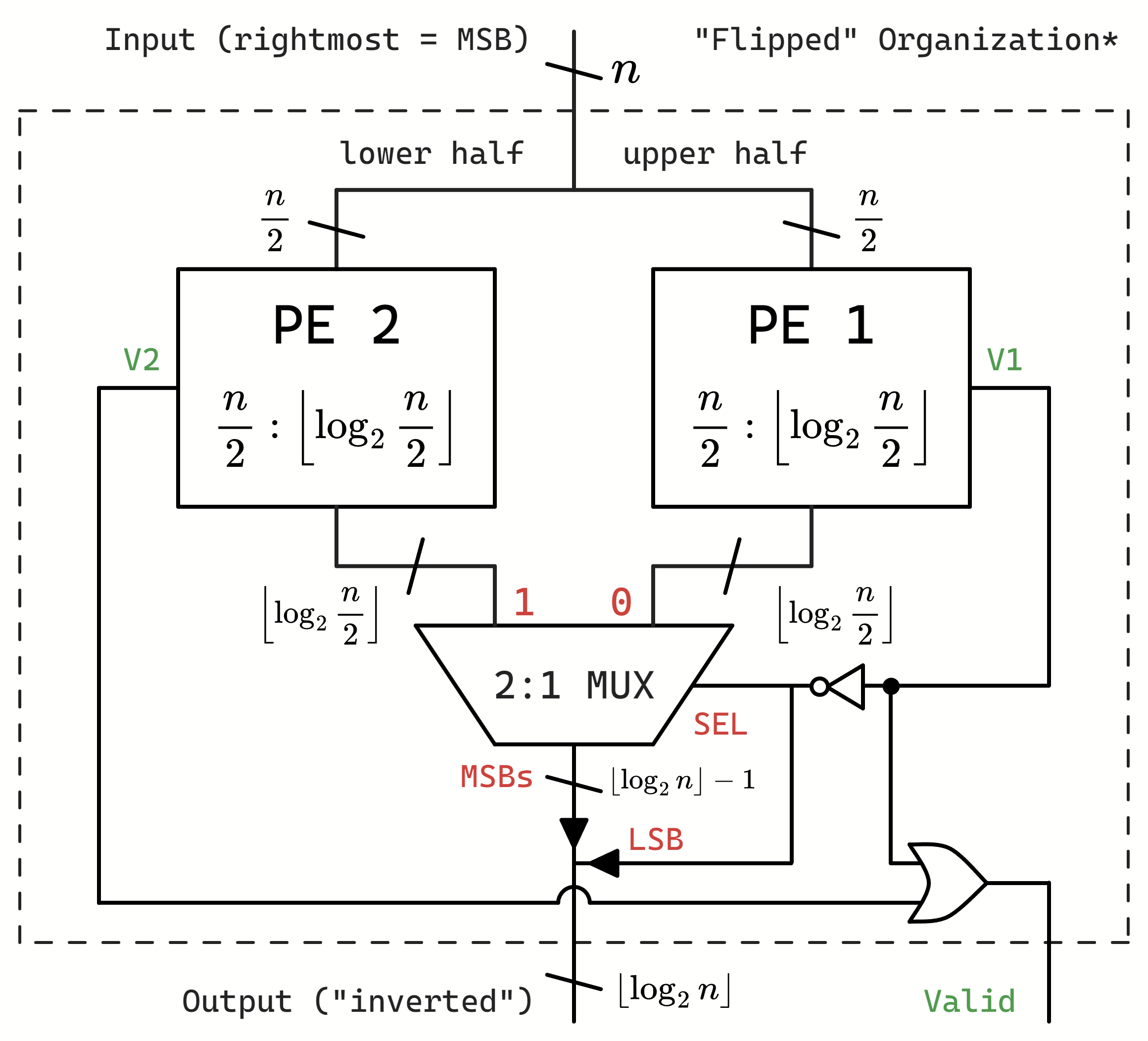}
    \vspace{-0.75em}
    \caption{Tree PE Structure, adapted from \cite{tree}. *This PE is organized opposite to all other PEs in this work.}
    \label{fig:tree_pe}
    \vspace{-0.75em}
\end{figure}

A tree-based priority encoder structure is introduced by \cite{tree}. 
This design is shown in Fig.~\ref{fig:tree_pe}.%
\footnote{Fig.~\ref{fig:tree_pe} is a reorganized but functionally equivalent version of the block diagram from \cite{tree}. This was done to unify notation throughout this work.}
Essentially, in this structure, the PE is recursively split in half with each layer. A $\lfloor \log_2n\rfloor$-bit 2:1 mux is used to select between the outputs of two sub-encoders.%
\footnote{In the original work \cite{tree}, the block diagram depicts a series of $\lfloor \log_2n\rfloor$ individual single-bit 2:1 muxes working in parallel. Since this is logically equivalent to a wider 2:1 mux, we have made this change in our diagram to enable easier comprehension.}
Critically, this PE design is organized \textbf{completely differently} than every other PE in this work. Specifically, the input is flipped such that the MSB is on the \textit{right} instead of the left. Thus, the output is essentially \q{inverted} compared to what our other PEs would produce; e.g., if the MSHB is the MSB, the output is $0$ rather than $2^n-1$. From a complexity and delay perspective, however, this fortunately makes no difference. Nevertheless, it is essential for understanding the following explanation.
The first/right PE takes the upper/most significant half of the input (i.e., bits $n-1$ down to $\frac{n}{2}$), and the second/left PE takes the lower/least significant half (bits $\frac{n}{2}-1$ down to $0$). If the output of PE 1 is valid, it is selected; otherwise, the output of PE 2 is selected. The valid output of PE 1 is concatenated with the selected sub-encoder output (i.e., as shown in Fig.~\ref{fig:tree_pe}, V1 is the LSB) to form the data output for the current layer of recursion. The valid output of the layer is the logical OR of the valid signals of both sub-encoders (i.e., V1 \code{OR} V2).

\begin{figure}[ht]
    \centering
    \vspace{-1em}
    \includegraphics[width=0.4\columnwidth]{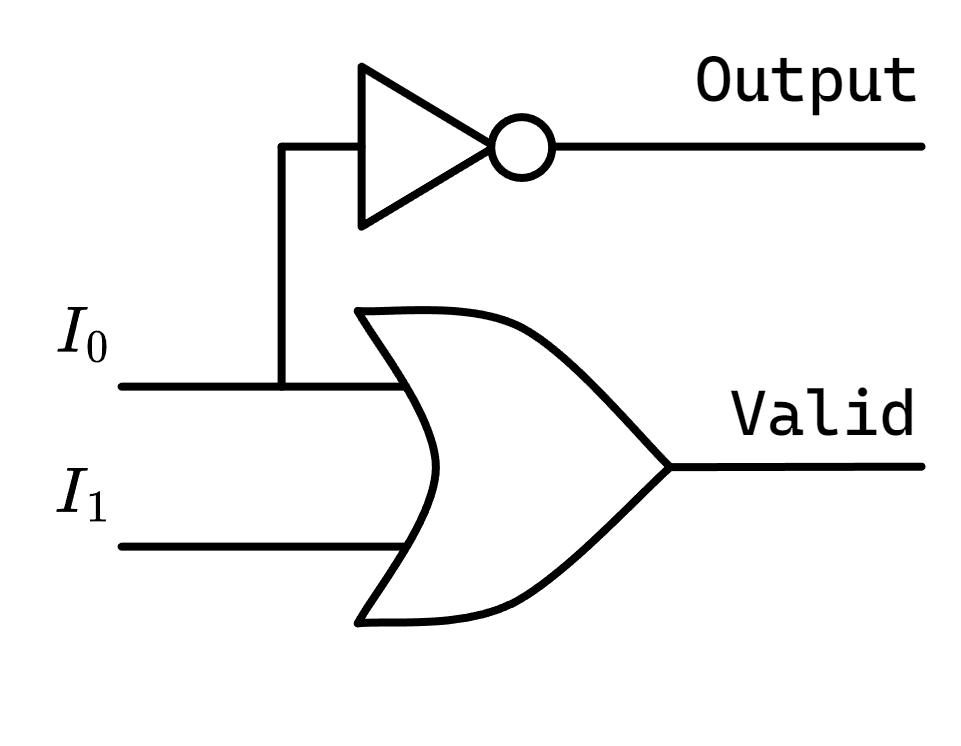}
    \vspace{-1.25em}
    \caption{Base case 2:1 PE (\q{flipped}) for tree structure, adapted from \cite{tree}.}
    \label{fig:tree_pe_base}
\end{figure}

In \cite{tree}, the base unit is a 2:1 PE consisting of a NOT gate and an OR gate (for the valid signal), as depicted in Fig.~\ref{fig:tree_pe_base}. Again, crucially, this PE is \q{flipped} such that $I_0$ is the MSB. For example, if $I_0=1$, then the output is $1$; if $I_0=0$ and $I_1=1$, then the output is $0$.%

It turns out that using $k=2$ for the recursive PE described in Section~\ref{sec:recursive_pe} is logically equivalent to this tree PE. To elaborate further, if for an arbitrary layer of the \q{tree} structure, we consider the NOT gate and OR gate to be a \q{flipped} 2:1 PE (as in Fig.~\ref{fig:tree_pe_base}), we find that the inputs are the valid signals of the sub-encoders, and the output is the select signal of the mux. Furthermore, for the entire encoder, the data output is the data output of the 2:1 PE concatenated with the output of the mux, and the valid output is the valid output of the 2:1 PE. This is effectively the same as the \q{recursive} design, as shown in Fig.~\ref{fig:recursive}, save for the organizational flip.

In this work, to align with the titles of the original works, we refer to the design shown in Fig.~\ref{fig:tree_pe} (or Fig.~\ref{fig:recursive} where $k=2$) as the \q{tree} PE design, and we refer to the design shown in Fig.~\ref{fig:recursive} where $k=4$ as the \q{recursive} PE design.

\subsection{Sequential Priority Encoders}\label{sec:seq_pe}
A discussion on priority encoders would not be complete without mention of sequential (or serial) priority encoders. All other architectures discussed in this paper operate in parallel and use exclusively combinational logic.%
\footnote{Technically, the original recursive design \cite{bcam} is registered between stages, but our recreation is not, so this statement effectively holds true.}
The benefits and drawbacks of serial vs. parallel architectures are generally well-known. Naturally, a serial PE will have a negligible complexity compared to any of the architectures discussed here. However, the delay of a serial PE will be correspondingly higher. For applications where it is paramount to reduce complexity at any cost (particularly without regard to delay), the serial PE is the obvious choice. This work primarily presents the multi-level structure, which is parallel, and compares it to similar designs. The goal of this work is not to compare serial and parallel architectures, but to assist designers in determining which parallel architecture is \q{best}---whatever that might mean for their particular use case---by providing a detailed analysis of complexity and delay for a variety of combinational parallel architectures.

\subsection{Other ASIC-Specific Methods}
Some works have implemented custom designs for PEs up to 256 bits through detailed changes at the transistor level \cite{chh1, chh2, chh3, dbnk}. Specifically, \cite{chh1, chh2, chh3} have implemented \q{multilevel lookahead and multilevel folding techniques.} This reduces the complexity, delay, and power substantially. \cite{dbnk} reproduces and further improves upon this design. However, these designs are essentially only valid for ASIC, i.e., transistor-level design.%
\footnote{While we analyze complexity, delay, and behavior at the transistor level for ASIC, our designs are functionally at the gate level (i.e., at a higher architectural level than these other works).}
Furthermore, our designs are primarily intended for higher precisions than 256 bits (the maximum considered by these works), and we also prefer to consider in general for both FPGA and ASIC. For these reasons, we did not re-implement the architectures proposed by these works.
Finally, note that \q{multilevel techniques} discussed in \cite{chh1, chh2, chh3, dbnk} are substantially different from the multi-level architectures proposed in this work.

\section{Two-Level Priority Encoders}\label{sec:2lpe}
\begin{figure}[ht]
    \centering
    \vspace{-0.75em}
    \includegraphics[width=\columnwidth]{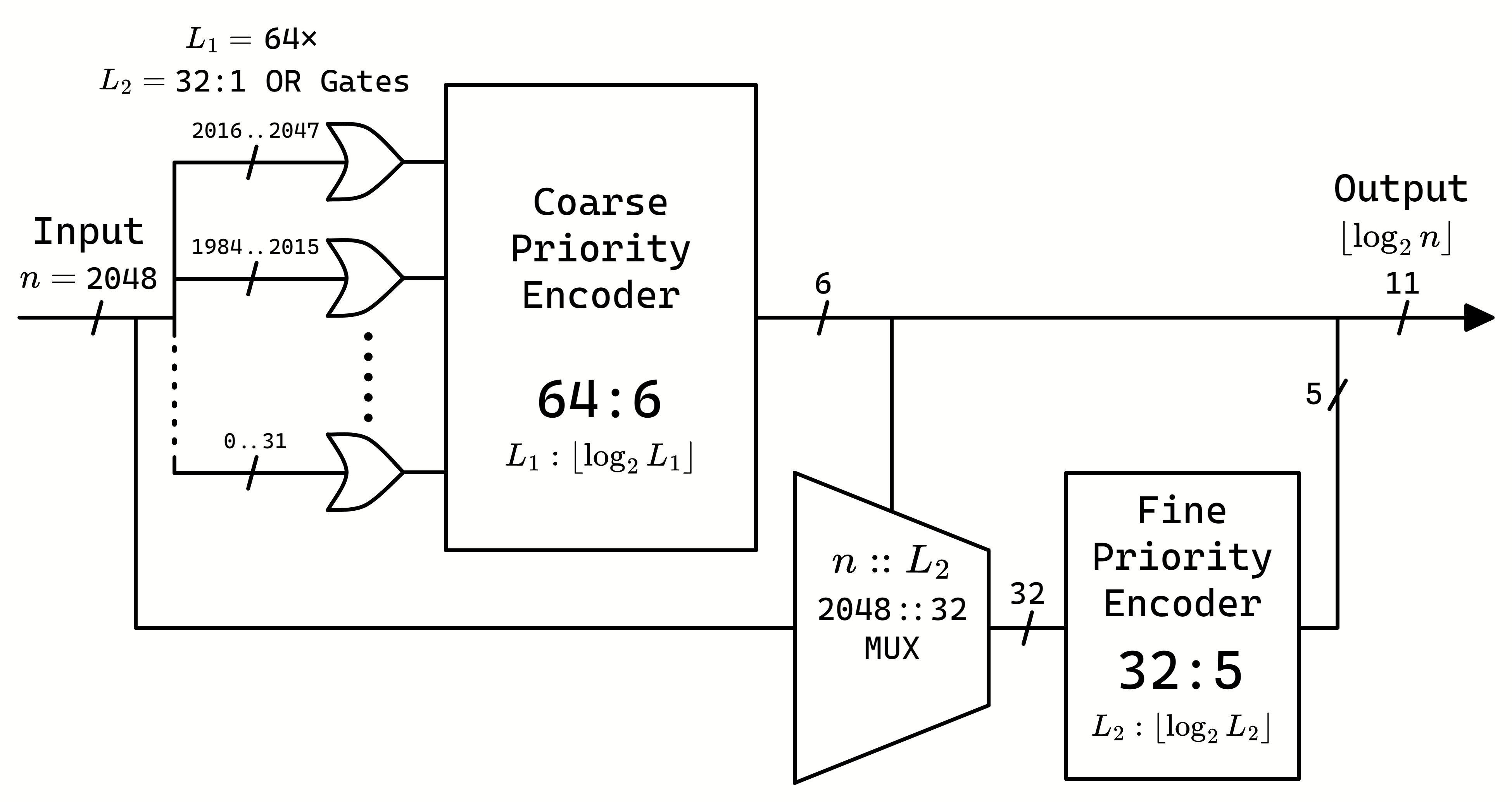}
    \vspace{-0.75em}
    \caption{A Block Diagram for a 2048-bit 2LPE, adapted from \cite{2lmr}.}
    \label{fig:2lpe-2048}
    \vspace{-0.25em}
\end{figure}

The idea originating this area of research was the \textit{two-level} priority encoder (2LPE). This hardware structure was independently developed by both our research group---introduced in \cite{2ldr} and further described by \cite{2lmr}---as well as the authors of \cite{xtn1} and \cite{xtn3}.\footnote{At the time of publishing \cite{2ldr} and \cite{2lmr}, the authors were not aware of \cite{xtn1} and \cite{xtn3}.} %
In this work, we build upon \cite{2ldr} and \cite{2lmr} with an improved description of the 2LPE compatible with our multi-level generalization. %
This two-level structure, shown in Fig.~\ref{fig:2lpe-2048}, is intended to address the problem of scaling at high input lengths.

First, we note that the designs and equations used to define the multi-level hardware (including two-level) in this paper are only guaranteed to work for input sizes that are a power of two. Furthermore, some special powers of two (perfect cubes or fourths) may result in more efficient hardware implementations depending on the level configuration.

Like an SLPE, a two-level priority encoder with $n$ inputs and $\lfloor \log_2n \rfloor$ outputs is denoted as an ${n:\lfloor \log_2n\rfloor}$ 2LPE. Unlike an SLPE, a 2LPE is constructed from several multi-input OR gates in parallel, a wide multiplexer, and two smaller PEs (denoted \textit{coarse} and \textit{fine}, respectively). The dimensions of each smaller PE are on the order of ${\sqrt{n}:\frac{1}{2}\log_2n}$. However, $n$ is not always a perfect square, so in order to keep the sizes of the sub-encoders as powers of two, we quantify them as follows.%
\footnote{In \cite{2lmr}, the input sizes of the coarse and fine were denoted $k$ and $q$, which are equal to $L_1$ and $L_2$, respectively.}
The input size $L_1$ of the coarse PE is defined by (\ref{eq:2lpe_coarse_param}). It is the smallest power of two such that $L_1\geq\sqrt{n}$:
\begin{equation}\label{eq:2lpe_coarse_param}
    L_1=2^{\displaystyle\big\lceil\log_{2}\sqrt{n}\,\big\rceil}.
\end{equation}
Meanwhile, the input size $L_2$ of the fine PE is defined by (\ref{eq:2lpe_fine_param}): 
\begin{equation}\label{eq:2lpe_fine_param}
    L_2=\frac{n}{L_1}.
\end{equation}
Then, referring to Fig.~\ref{fig:2lpe-2048}, the 2LPE functions as follows: 
\begin{enumerate}[label=\textbf{\arabic*.}]
    \item The input is divided into several \q{slices} by $L_1$ parallel OR gates of size ${L_2:1}$.
    \item The outputs of the OR gates form the input of the coarse PE, which finds the slice that the MSHB is located in.
    \item The output of the coarse PE becomes the select signal for the multiplexer, which is used to select the appropriate slice for the fine PE.
    \item The fine PE finds the exact position of the MSHB within that slice.
    \item The output of the entire 2LPE is created by concatenating the outputs of the coarse and fine PEs, in that order.
\end{enumerate}

Finally, note the unique notation for the wide multiplexer in Fig.~\ref{fig:2lpe-2048}. 
A simple multiplexer taking four bits as input and selecting one bit as output might be denoted a \q{$4:1$ mux} or \q{$4$-to-$1$ mux}.
Meanwhile, a mux consisting of four input channels and one output channel, where each channel is two bits wide, might fully be described as a \q{2-bit $4:1$ mux.}\footnote{Although, of course, this mux would be implemented as two of the aforementioned simple muxes, working in parallel.}
This type of multiplexer, with one wide output channel, is used extensively in our designs. To enable brevity without ambiguity, we introduce a new double-colon notation, $x::y$, where $x$ is the total number of input bits, and $y$ is the total number of output bits. All muxes denoted in this way have only one output channel; thus, $y$ is also the number of bits per channel. In other words, all channels are $y$ bits wide. The actual number of input channels is then $x/y$. For example, Fig.~\ref{fig:2lpe-2048} shows a $2048::32$ mux with $64$ input channels and one output channel, all of which are each $32$ bits wide.\footnote{Thus, this $2048\!::\!32$ mux might also be described as a 32-bit $64\!:\!1$ mux.} Finally, all muxes denoted as $x:1$ are single-bit with $x$ input channels and one bit/channel output. 

\section{Multi-Level Priority Encoders}\label{sec:mlpe}
We first establish the notation used to refer to multi-level priority encoders. First, in Section~\ref{sec:slpe}, we denote a standard PE as a single-level priority encoder (SLPE). As shown in Section~\ref{sec:2lpe}, we refer to a two-level priority encoder as a \q{2LPE.} \footnote{A two-level priority encoder was previously referred to as a \q{TLPE} in \cite{2lmr}, as that notation was created before generalizing to multiple levels was considered. To eliminate confusion with \q{\underline{t}hree}, we now use \q{2LPE.}}
Similarly, a three-level priority encoder is a \q{3LPE,} and so on. \q{MLPE} refers to an \q{$m$-level} or \q{multi-level} priority encoder. That is, a PE with an arbitrary number of levels $m$, where $m\in \{2,3,4,...\}$. 
Next, we discuss the two different methods for constructing MLPEs---\textit{composition} and \textit{cascading}---with a focus on the three-level case.

\subsection{Composition}\label{sec:mlpe-composition}
\begin{figure}[h!]
    \centering
    \vspace{-0.75em}
    \includegraphics[width=0.86\columnwidth]{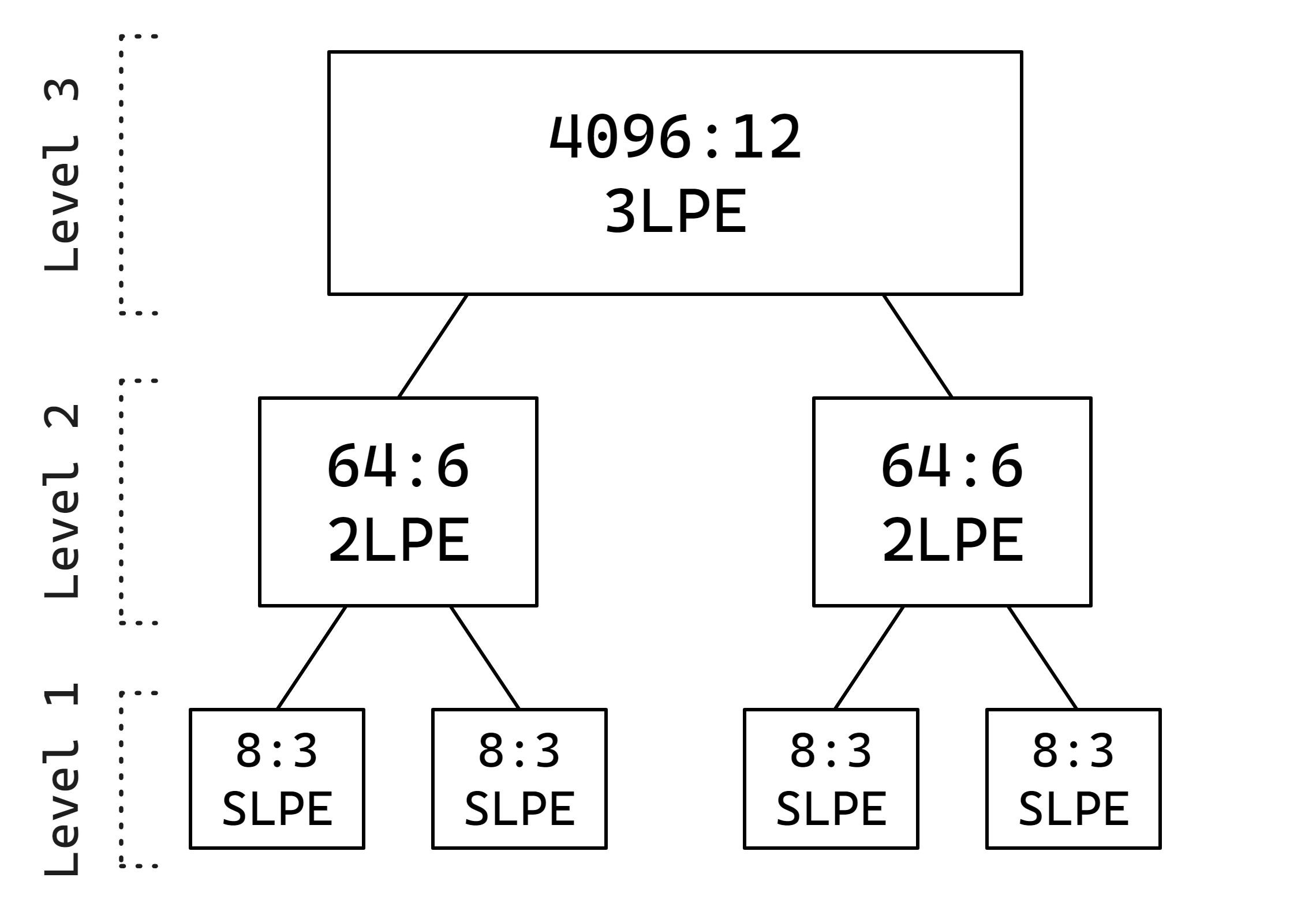}
    \vspace{-0.75em}
    \caption{Tree Diagram for a 4096-bit 3LPE.}
    \label{fig:4096-3lpe-comp-tree}
    \vspace{-1em}
\end{figure}

The first method, \q{composition,} was derived from a programmatical or hardware description perspective. When testing large 2LPEs, the sub-encoders became large enough that we hypothesized that complexity could be reduced by making the sub-encoders themselves 2LPEs. As the smaller encoders were originally separate VHDL components, it was trivial to replace those components with 2LPEs of an identical size. To clarify our definition of the number of levels $m$ for this paradigm, it is easiest to view this encoder as a tree, as shown in Fig.~\ref{fig:4096-3lpe-comp-tree}.

\begin{figure*}[t!]
    \centering
    \includegraphics[width=\textwidth]{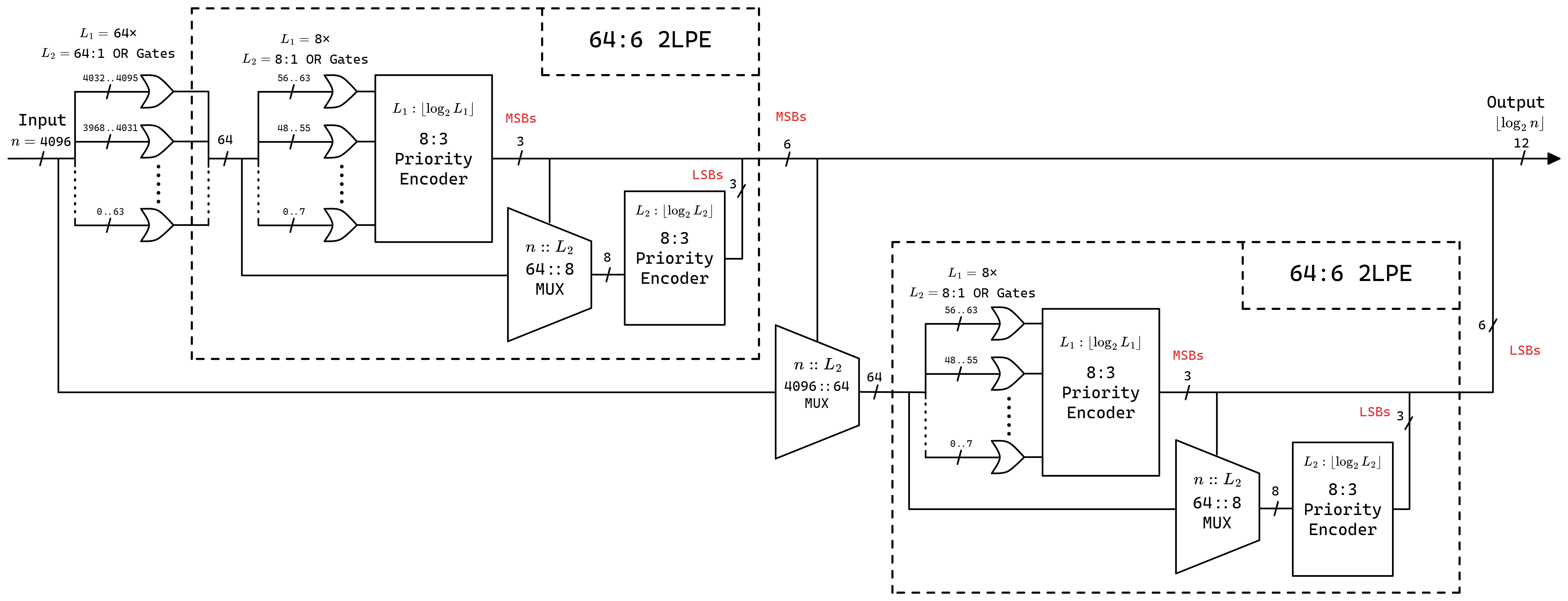}
    \vspace{-0.75em}
    \caption{A block diagram for a 4096-bit composed 3LPE.}
    \label{fig:4096-3lpe-comp-diag}
\end{figure*}

Based on Fig.~\ref{fig:4096-3lpe-comp-tree}, it is clear that this encoder has three levels, and as such is called a 3LPE. We note that \textit{composed} MLPEs are simplest to construct (or most balanced) when the input length $n$ is a perfect square, or better yet, a perfect fourth (power) such as 4096. For example, Fig.~\ref{fig:4096-3lpe-comp-diag} is a block diagram showing a 4096:12 composed 3LPE. The dotted outlines each enclose a 64:6 2LPE.

\begin{figure*}[ht!]
    \centering
    \includegraphics[width=\textwidth]{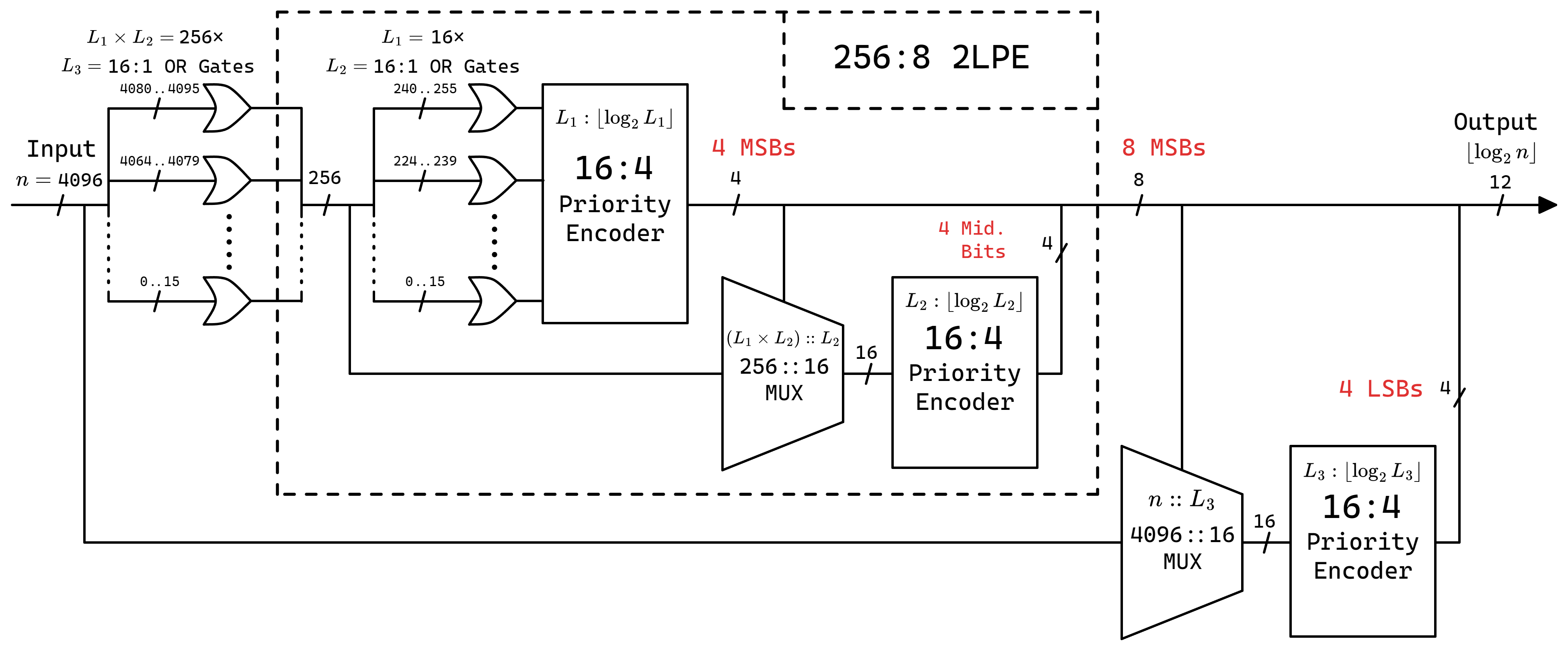}
    \vspace{-0.75em}
    \caption{A block diagram for a 4096-bit cascaded 3LPE.}
    \label{fig:4096-3lpe-cascade-diag}
\end{figure*}

\subsection{Cascading}\label{sec:mlpe-cascading}
The second method, derived from a more theoretical perspective, is \q{cascading} or \q{chaining.} The easiest way to visualize this method is by assuming that the coarse encoder in a standard 2LPE is itself replaced by another 2LPE. This requires changes to how the size parameters of each encoder are defined. %
Critically, $L_i$ for cascaded MLPEs are \textit{not} defined the same as composed MLPEs. In the composed structure, each sub-encoder effectively has its own $L_1$ and $L_2$, where in the cascaded structure, $L_i$ are unified across all levels. Thus, we redefine $L_i$ for a cascaded MLPE of $m$ levels such that $1\leq i \leq m$, and $\prod_{1}^{i}L_i=n$, where $L_i$ are now defined by (\ref{eq:mlpe_cascade_L_1}) for $i=1$ and (\ref{eq:mlpe_cascade_L_i}) for $i>1$:
\begin{equation}\label{eq:mlpe_cascade_L_1}
L_1=2^{\displaystyle\big\lceil\log_{2}\sqrt[m]{n}\,\big\rceil}, 
\end{equation}
\begin{equation}\label{eq:mlpe_cascade_L_i}
L_{i}=2^{\displaystyle \left\lceil\log_{2}\left(\left(n \cdot \prod_{j=1}^{i-1}\frac{1}{L_{j}}\right)^\frac{1}{m-i+1}\right)\right\rceil}.
\end{equation}
As an example: for a cascaded 3LPE, $L_1$ is the least power of $2$ greater than or equal to $\sqrt[3] n$, or $2^{\lceil\,\log_2\sqrt[3] n\,\rceil}$. $L_2$ is the least power of $2$ greater than or equal to $\sqrt{n / L_1}$. $L_3$ is $(n/L_1)/L_2$. Typically, $L_1=L_2$. If $n$ is a perfect cube, $L_1=L_2=L_3=\sqrt[3]{n}$. It follows that with cascading for $m=3$, perfect cubes are the simplest to use, as opposed to perfect fourths for composition or $m=4$ cascading. An example of a 4096:12 cascaded 3LPE is shown in Fig.~\ref{fig:4096-3lpe-cascade-diag} on the next page.

To elaborate on the function of (\ref{eq:mlpe_cascade_L_i}), which is rather obtuse at first glance: fundamentally, the goal of these equations is to derive the sizes of the components in a generalized manner. Since we are working with powers of two, we have to ensure that all component sizes conform to this requirement. The outside component, $2^{\lceil\log_2(...)\rceil}$ accomplishes this. This represents \q{the least power of 2 greater than $x$.} We also want to have balanced and reasonable sizes, not arbitrary ones, and the sizes must be whole numbers. For convenient cases like $m=3$ and $n=4096$, it may seem straightforward to choose all sizes to be $\sqrt[m]{n}=\sqrt[3]{4096}=16$. However, when $n$ is not a perfect cube (etc.), this would result in component sizes that are not whole numbers, much less powers of two. Thus, we successively divide $n$ by each preceding size, guaranteed to be a power of two. Then we take a root of this quantity, for which the fractional exponent decrements as $i$ increases, starting at $m$ for $i=1$. In fact, (\ref{eq:mlpe_cascade_L_1}) is a logically simplified version of (\ref{eq:mlpe_cascade_L_i}). However, whether $i=1$ actually works with (\ref{eq:mlpe_cascade_L_i}) depends on how the calculator or computer handles a product where the initial iterator is greater than the limit. Using a distinct base case, i.e., (\ref{eq:mlpe_cascade_L_1}), eliminates the possibility for ambiguity or implementation-induced error.

\subsection{Valid Signal}\label{sec:mlpe-valid-sig}
Regardless of the method of construction (composition or cascading), adding a \textit{valid} signal to MLPEs is trivial, thanks to the pre-existing OR gates connected to all $n$ inputs. 
The straightforward method is to add an additional OR gate after the first stage, before the coarse or $L_1$ encoder. %
This method was used for the FPGA results described in Section~\ref{sec:results}. %
However, we can take advantage of the idea that the purpose of the valid signal can be expressed as \q{to differentiate between an input of all zeroes and an input of one (\code{00...01}).} Then, we can simply OR the least significant bit of the input with the entire output, which can be less complex with large $n$.

\vspace{0.95em}

\subsection{Differences from Multi-Level Decoders}\label{sec:mlpe-vs-mld}
In \cite{mld}, we conducted a similar analysis of the multi-level paradigm applied to \textit{decoders} instead of priority encoders. In this work, we also analyze various architectures, compare their complexity and delay, and provide usage recommendations. However, our analysis here is much deeper, in part due to the increased complexity of PEs.
Furthermore, the results differ substantially from \cite{mld}, as further described in Section~\ref{sec:results}. Additionally, MLPEs themselves exhibit some key differences compared to MLDs, aside from the primary difference in functionality. In particular, the multi-dimensional approach to constructing MLDs is not possible with MLPEs. 

\section{Components and fan-in}\label{sec:components}
First, we define an \textit{atomic} or \textit{base} component, such as a logic gate or PE, as one instance of that component that is not divisible into or constructed out of other discrete instances of that component. For example, consider a binary tree consisting of seven 2-input OR gates, each implemented directly out of six CMOS transistors in the typical fashion. When viewed as a unit, they function as an 8-input OR gate. Individually, they can be considered \textit{atomic} components, but the gestalt circuit is not atomic. Furthermore, we define a component containing multiple atomic components, such as this example, to be \textit{composite}.

When considering an abstract, composite gate with an arbitrary number of inputs, defining at what level the component gates should be implemented atomically is a key design criterion. This decision is closely linked with the concept of \textit{fan-in}.
Careful consideration of these two intertwined factors is particularly crucial for high-precision priority encoders and their components, since they have many inputs.
\footnote{Fortunately, since PEs encode many inputs into few outputs, internal fan\textit{out} is not a concern.} %
Weste and Harris state that \q{it is rarely advisable to construct a gate with more than four or possibly five series transistors} \cite{vlsi}. Since each additional input requires another transistor in series, and input sizes corresponding to powers of two are most convenient, the maximum practical fan-in for atomic logic gates and multiplexers is four \cite{vlsi, le}.

The two aforementioned elements are the basic building blocks of multi-level priority encoders.
MLPEs consist of large OR gates, wide multiplexers, and smaller PEs (or sub-encoders). The latter are also composed of logic gates, whether 
as used in multiplexers, or when implemented directly according to their boolean equations. In order to optimize delay and complexity, we make several observations---and consequently, decisions---about these components.
\subsection{OR Gates}\label{sec:components:or_gates}

Moving forward, except where otherwise stated, we assume that any gate of four or fewer inputs is an atomic gate, and that any gate with more than four inputs is implemented as a composite structure of atomic gates. This aligns with our established maximum fan-in of four. Additionally, we will use a shorthand notation concatenating the gate name and number of inputs. For example, \q{OR8} refers to an 8-input OR gate.

\begin{figure}[ht]
    \centering
    \vspace{-0.75em}
    \includegraphics[width=0.5\columnwidth]{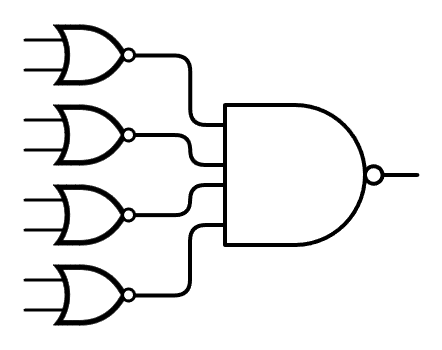}
    \vspace{-0.75em}
    \caption{Atomic components of OR8 unit.}
    \label{fig:or8u}
    \vspace{-0.25em}
\end{figure}

As shown in Fig.~\ref{fig:or8u}, we chose to construct OR gates of width 8 bits or larger out of the OR8 \q{unit} structure in order to optimize delay (or speed) and complexity (or area, and by extension, power).

Assuming a maximum fan-in of four for the base gates as previously established, we can use atomic NAND4 and NOR4 gates. Again, for convenience, we generally only consider input sizes corresponding to powers of two. 
Thus, we do not require or use 3-input atomic gates in nearly all cases,\footnote{Except for multiplexers, as discussed in \ref{sec:components:muxes}.} although they could be helpful to optimize for use cases where the input size is not a power of two.

Furthermore, the usage of four NOR2 gates feeding a NAND4 gate is deliberate. In general, NOR gates of an arbitrary input size $n$ are slower than $n$-input NAND gates because NOR gates have $n$ PMOS transistors in series, which are slower than NMOS transistors. Meanwhile, NAND gates have the faster NMOS transistors in series and the slower PMOS transistors in parallel \cite{vlsi}.
Thus, we chose to minimize the size and usage of NOR gates in our OR trees by maximizing the size and usage of NAND gates.

In total, this OR8 unit requires 24 transistors, which is only 4 transistors more than the minimal-complexity\footnote{That is, using gates with at most 4 inputs.} implementation with two NOR4 gates feeding a NAND2 gate. In cases where minimizing complexity is especially important, the aforementioned 20-transistor version---or even an atomic 18-transistor version---can be used, but the reduction in speed may quickly become noticeable with many composite OR gates in series within larger priority encoders.

Next, we will define a composite OR4 unit as two NOR2 gates feeding a NAND2 gate, similar to the OR8 unit. This requires 12 transistors, which is only 2 more than an atomic OR4 gate, and has fewer PMOS transistors in series.
Now, we can utilize the standard OR2 gate (NOR followed by an inverter) along with the aforementioned OR4 and OR8 units to create composite OR gates of arbitrary size. By creating a \textit{leaf-heavy}%
    \footnote{By \textit{leaf-heavy} we mean that the size of the OR units closest to the inputs (leaves) should be maximized, while the size of the OR units closest to the output (root) should be minimized. Note that this is the opposite of the philosophy behind the OR8 unit.} 
tree of the various OR units, we can strike a balance between complexity and delay. %

\subsection{Multiplexers}\label{sec:components:muxes}
Comparatively speaking, our analysis of multiplexers is much more straightforward. Sutherland \textit{et al.} claim that \q{it is best to partition wide multiplexers into trees of four-input multiplexers} \cite{le}, which also aligns with our previously established fan-in limit of four inputs \cite{vlsi, le}.
Consider the optimal implementation of a four-input mux, shown in Fig.~\ref{fig:mux4} below, alongside our analysis of logic gates.

\begin{figure}[ht]
    \centering
    \includegraphics[width=0.7\columnwidth]{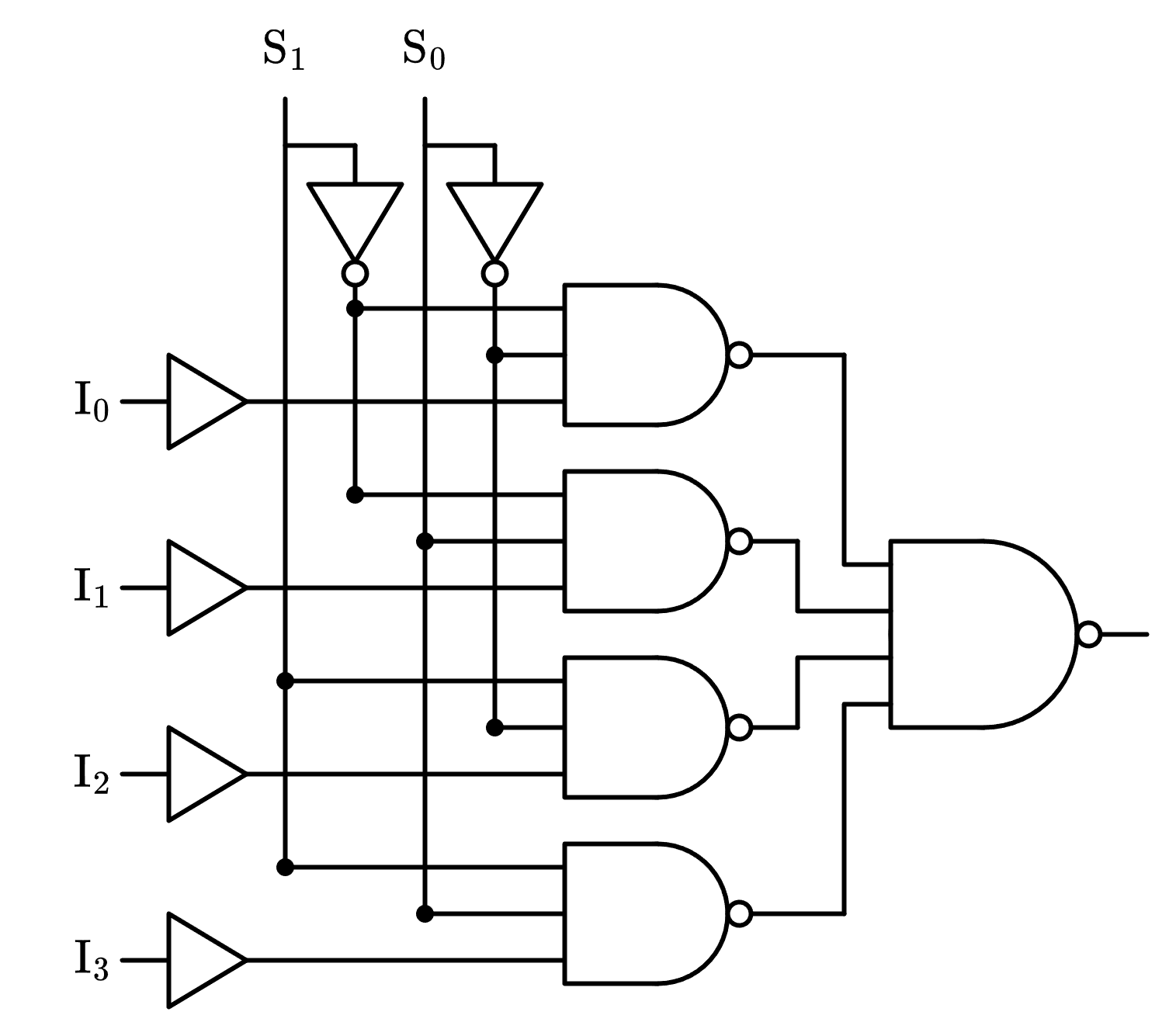}
    \vspace{-0.75em}
    \caption{Gate-level implementation of four-input, single-bit multiplexer.}
    \label{fig:mux4}
    \vspace{-0.75em}
\end{figure}

\begin{figure}[ht]
    \centering
    \vspace{-0.75em}
    \includegraphics[width=0.3\columnwidth]{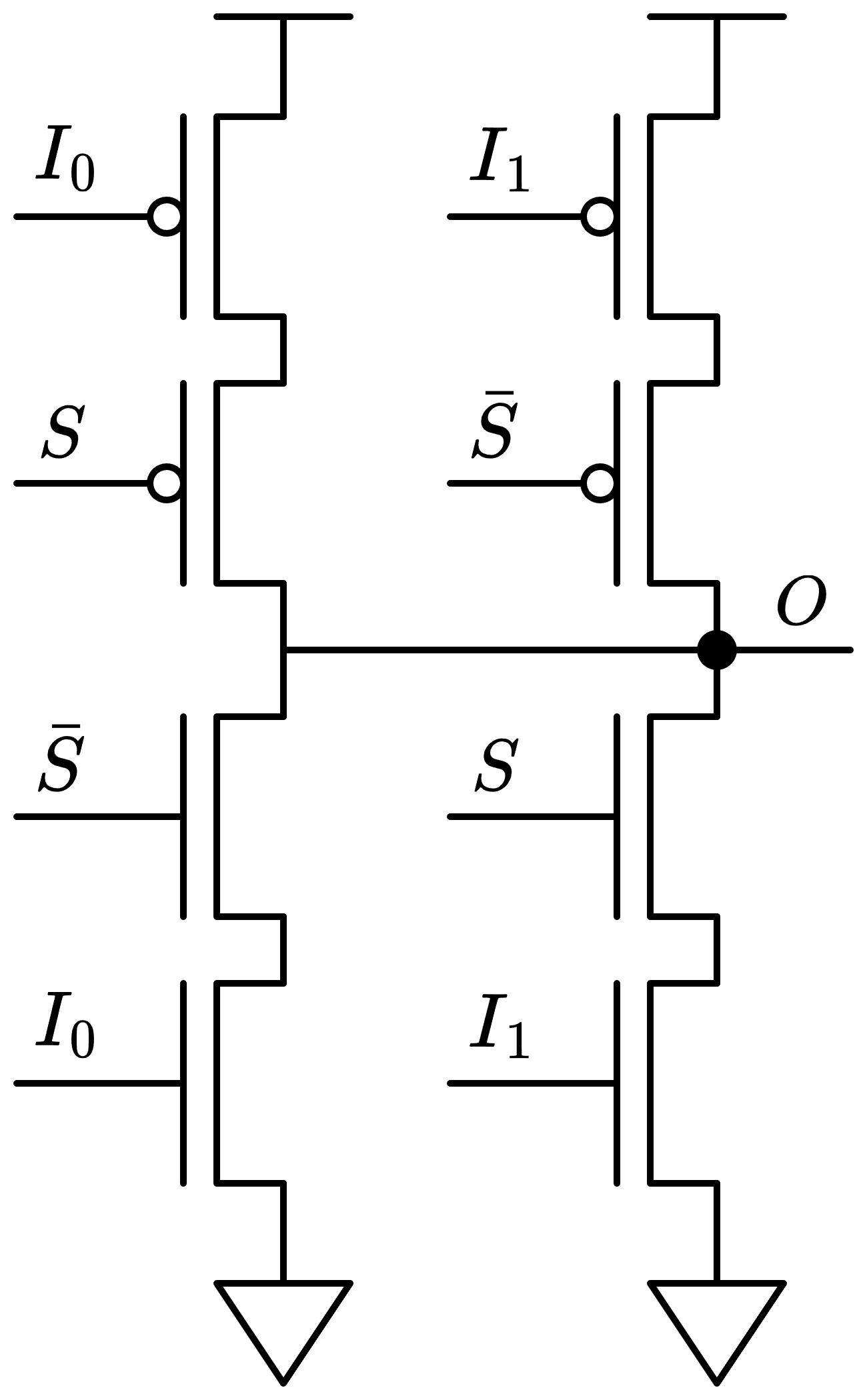}
    \vspace{-0.75em}
    \caption{Transistor-level implementation of a 2::1 mux, adapted from \cite{vlsi}.}
    \label{fig:mux2a}
    \vspace{-0.25em}
\end{figure}

In Fig.~\ref{fig:mux4}, we can see that a four-input mux can be implemented using four NAND3 gates feeding a NAND4 gate.
Including the two inverters (one for each bit of the select signal), we obtain a total cost of 36 CMOS transistors (8 for the NAND4, 6 for each NAND3, and 2 for each inverter).
Thanks to De Morgan's Laws, no NOR gates need to be used here. This is the fastest and cheapest way to achieve this functionality \textit{at the gate-level}.
We can do better if we can use single-bit 2:1 muxes composed directly from 8 static CMOS transistors, as shown in Fig.~\ref{fig:mux2a}, instead of one composed from logic gates.

From three of these, we can compose a 4:1 mux for only 24 transistors, and, as shown in Section~\ref{sec:methods-asic2}, with a slightly lower delay than the previously described gate-level method.

\subsection{Atomic Priority Encoders}\label{sec:components:atomic_pes}
A central idea of this work is the benefits achieved by dividing a large priority encoder into several smaller priority encoders. The primary novel contribution of this work is generalizing this idea to multiple levels. At the lowest level lies the atomic priority encoders, which follow the previous definition of \textit{atomic}.
Atomic PEs are SLPEs, thus, they can be implemented in the two ways described in Section~\ref{sec:slpe}: either by using gates directly or by using a chain of muxes.

\subsection{FPGA Components}\label{sec:components:fpga}
The components of FPGA development boards are well known, but they are useful for our PE designs in a few important ways, which we review here. The development boards we work with, based on the AMD Artix-7 FPGA, contain many Configurable Logic Blocks (CLBs), the main organizational unit of the device. Each CLB contains two logic slices, with each slice containing 4 lookup tables (LUTs) and 8 flip-flops (FFs) \cite{ug474}.
These implementation details are handled by the synthesis tool, but the important detail for us is that the availability of LUT6 primitives makes it trivial to implement a $4::1$ mux, as mentioned in Section~\ref{sec:recursive_pe}.
\begin{figure}[ht]
    \centering
    \includegraphics[width=0.9\columnwidth]{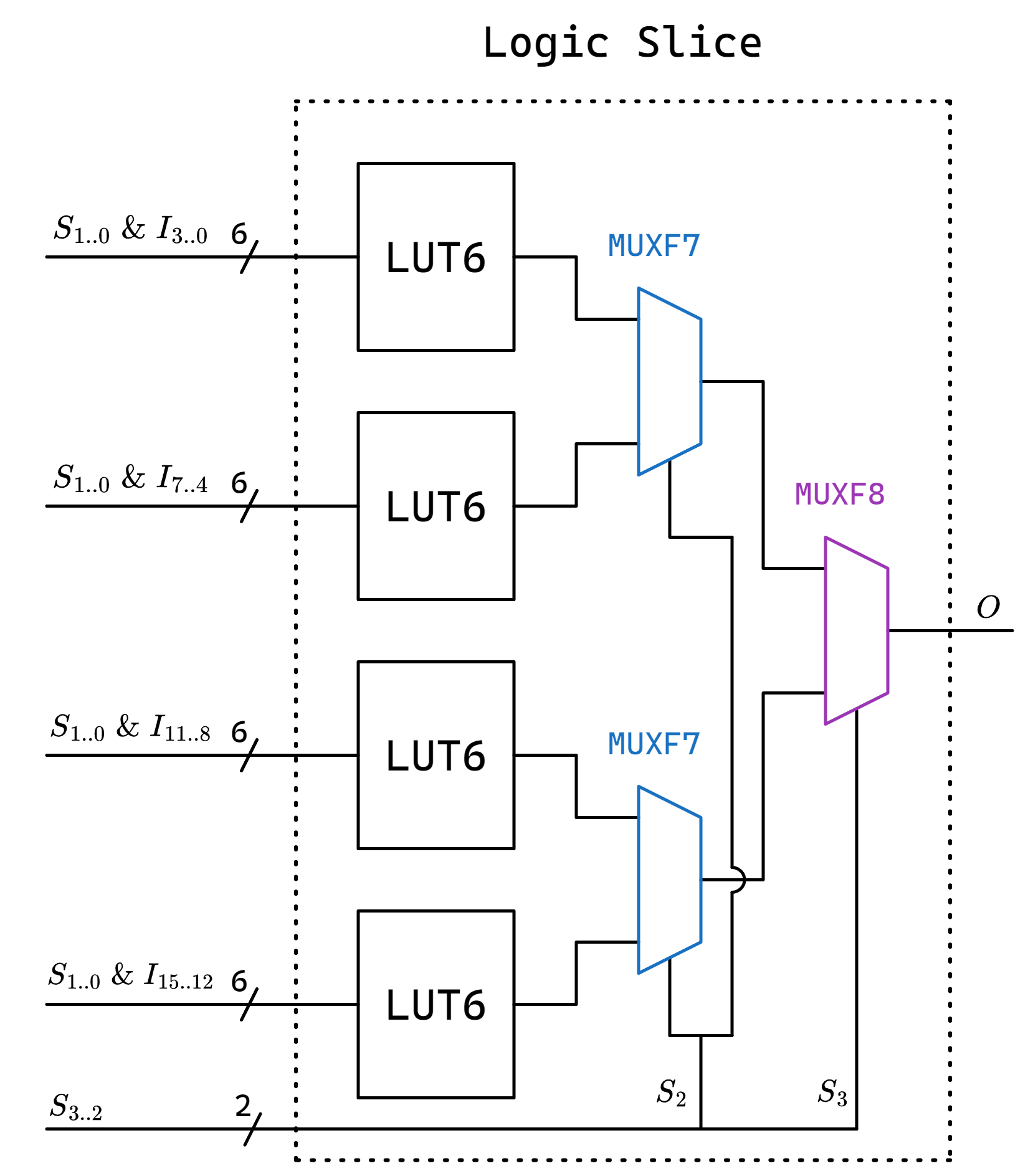}
    \vspace{-1em}
    \caption{A simplified Artix-7 FPGA logic slice, configured as a 16::1 mux, adapted from \cite{ug474}. The symbol \code{\&} denotes concatenation.}
    \label{fig:logic-slice}
    \vspace{-0.75em}
\end{figure}

Furthermore, as shown in Fig.~\ref{fig:logic-slice}, in addition to the four LUT6 primitives, each logic slice contains two MUXF7 primitives and one MUXF8 primitive (collectively, \q{MUXFX primitives}). These components allow easy implementation of wide multiplexers -- either four $4::1$, two $8::1$, or one $16::1$ mux(es) on a single logic slice. Thanks to this capability, the large muxes used in our MLPEs are substantially easier to construct on an FPGA than on an ASIC, since in the latter technology, our atomic muxes are at most $4::1$, as detailed in Section~\ref{sec:components:muxes}. We will see the effects of this in Section~\ref{sec:results}. 

\section{Methodology}\label{sec:methods}
\subsection{FPGA Synthesis}\label{sec:methods-fpga}
We synthesized several PE designs described in VHDL with AMD Vivado Design Suite for a Digilent Nexys-A7 100T development board. We compare complexity (in total LUT usage) and delay (in LUTs on the critical path) for the standard (SL)PE, our 2LPE, three and four level MLPEs (both composed and cascading), and the recursive method from \cite{bcam}. Based on our findings from \cite{mld} and our initial experimentation, we decided not to collect detailed results for MLPEs where $m\geq5$ -- five levels is excessive for the input lengths under consideration here, and as discussed in Section~\ref{sec:results-dim-ret}, does not provide any additional reductions in either complexity or delay for those input lengths. Furthermore, for the atomic encoders, we chose to report results using mux-based SLPEs%
\footnote{Refer to Section~\ref{sec:slpe} for the differences between mux-based and gate-based SLPEs.}
as the base encoders, since our initial testing revealed that---except for a few negligible exceptions---using muxes is always cheaper for MLPEs.

The full VHDL descriptions are available in our GitHub repository at \href{https://github.com/ALUminaries/Multi-Level-Priority-Encoders}{github.com/ALUminaries/Multi-Level-Priority-Encoders}. To summarize, the main VHDL file is capable of generating any reasonable size of MLPE (cascaded or composed) or SLPE (mux-based or gate-based). The exact hardware generated is determined via the generics. The input and output sizes of the PE are \code{G\_n} ($n$) and \code{G\_lg\_n} ($\log_2 n$), respectively. The architecture of the PE is determined by \code{G\_use\_gate\_optimized} (\code{true} to use gate-based SLPEs, \code{false} to use mux-based) and \code{G\_use\_cascading} (\code{true} to generate a cascading MLPE, \code{false} to generate a composed MLPE). Note that for $m\leq 2$, \code{G\_use\_cascading} has no effect, as in this case the hardware generated is either a 2LPE (generated using the composed path) or an SLPE. The method of determining the actual number of levels, $m$, is slightly more complex to account for edge cases, but in most cases, $m$ is equal to the configured value for \code{G\_max\_lvls}.%
\footnote{The relevant generic, \code{G\_max\_lvls}, is technically not $m$, but rather a \textit{limit} on $m$. Depending on the configured values of \code{G\_n}, \code{G\_max\_lvls}. and \code{G\_use\_gate\_optimized}, the generated hardware may not reach $m=$~\code{G\_max\_lvls}, but will never exceed \code{G\_max\_lvls}.} Also note that all synthesized hardware includes valid signals, as described in Section~\ref{sec:mlpe-valid-sig}.

To record the FPGA results discussed in Section~\ref{sec:results-fpga}, we ran Vivado synthesis with the default configuration, targeted at the \code{xc7a100tcsg324-1} device. To determine complexity, we recorded the LUT utilization as reported by the synthesized design interface, as well as the number of MUXFX primitives used as reported by the \code{report\_utilization} command. Finally, in order to normalize complexity into a single data value and allow more fair comparisons between those architectures using the MUXFX primitives and those without, we converted the MUXFX usage to normalized LUT equivalents ($\text{LUT}_\text{N}$) using \eqref{eq:muxfx_lut_conv}. 
\begin{equation}\label{eq:muxfx_lut_conv}
\text{LUT}_\text{N} = \text{LUT} + \left\lceil\frac{\text{MUXFX}}{3}\right\rceil.
\end{equation}
This conversion is based on the fact that a single LUT6 can implement a $4::1$ mux, which itself is composed of three $2::1$ muxes, each of which would be equivalent to a single MUXFX primitive.%
\footnote{In other words, because a MUXFX primitive is a $2::1$ mux, we translated MUXFX to LUTs at a ratio of approximately 3:1.}

Meanwhile, to determine the delay, we summarized results from the \code{report\_timing} command. We report the total number of components (LUTs and MUXFX primitives) along the critical path. Following the same logic as for complexity, MUXFX are counted at half the value of a LUT.%
\footnote{Since the critical path of a $4::1$ mux would be two $2::1$ muxes, and we consider a $4::1$ mux equivalent to a LUT, a single $2::1$ mux should have half the delay as passing through two.}

Furthermore, for the purposes of FPGA delay, we considered the delay of a single LUT to be constant regardless of how many inputs Vivado reported it as having. This is for two reasons: 1. Every LUT on the FPGA is essentially identical, excepting manufacturing tolerances, and 2. The official documentation states that "the propagation delay through a LUT is independent of the function implemented" \cite{ug474}. It is also worth noting that in modern FPGAs, it is well-known that the majority of delay is due to routing, rather than actual propagation delay. However, since routing delay is subject to many more variables (device, run-to-run variance, etc.), we chose to report the critical path in terms of LUTs, largely because it is used for relative comparisons among the architectures tested, rather than a statement of absolute speed. Using this and other techniques, we provide a more general overview and comparison that may prove more useful to various users and applications.

\subsection{ASIC Complexity Analysis}\label{sec:methods-asic}
In this section, we derive formulas that can be used to calculate the complexity, in static CMOS transistors, of the various PE designs when implemented on an arbitrary ASIC. Furthermore, we provide a Desmos calculator that may be used to evaluate and visualize the equations in this section here: \href{https://www.desmos.com/calculator/nui3qzcfhs}{www.desmos.com/calculator/nui3qzcfhs}.%
\footnote{This link is up-to-date at time of writing; if updated, the new link will be available on our GitHub page from Section~\ref{sec:methods-fpga}.}
Additionally, note that the complexity/cost equations given here (and thus the results discussed in Section~\ref{sec:results-asic}) do not account for the transistors required to generate a valid signal for any architectures considered.

\subsubsection{Base Logic Gate Costs}
We first establish expressions for the cost (or complexity) in terms of the number of transistors, denoted by $C$, for the various basic logic gates utilized in PEs (subscripted). We define these using standard static CMOS technology; as a result, the equations in the following subsections may be used for any ASIC technology, provided these base costs are adjusted appropriately. This can be accomplished easily by changing the appropriate variables in the aforementioned Desmos calculator. 

Beginning with the simplest gates, the cost in transistors of the NOT gate (or inverter), the 2-input NAND gate, and the 2-input NOR gates are represented by \eqref{eq:C_NOT}, \eqref{eq:C_NAND2}, and \eqref{eq:C_NOR2}, respectively \cite{vlsi}.\footnote{We use \textit{CMOS VLSI Design} \cite{vlsi} as a reputable cross-reference for gate schematics at the transistor level.}
\begin{equation}\label{eq:C_NOT}
\cost{NOT}=2.
\end{equation}
\begin{equation}\label{eq:C_NAND2}
\cost{NAND2}=4.
\end{equation}
\begin{equation}\label{eq:C_NOR2}
\cost{NOR2}=4.
\end{equation}

Equations \eqref{eq:C_AND2} and \eqref{eq:C_OR2} represent the 2-input AND and OR gates, respectively:
\begin{equation}\label{eq:C_AND2}
\cost{AND2}=\cost{NAND2}+\cost{NOT},
\end{equation}
\begin{equation}\label{eq:C_OR2}
\cost{OR2}=\cost{NOR2}+\cost{NOT}.
\end{equation}

The following equations represent the atomic NAND3 gate \eqref{eq:C_NAND3A} and the atomic 4-input gates for NAND \eqref{eq:C_NAND4A}, NOR \eqref{eq:C_NOR4A}, and OR \eqref{eq:C_OR4A} \cite{vlsi}. For clarity, the `A' suffix of the subscript denotes an atomic gate.
\begin{equation}\label{eq:C_NAND3A}
\cost{NAND3A}=6.
\end{equation}
\begin{equation}\label{eq:C_NAND4A}
\cost{NAND4A}=8
\end{equation}
\begin{equation}\label{eq:C_NOR4A}
\cost{NOR4A}=8.
\end{equation}
\begin{equation}\label{eq:C_OR4A}
\cost{OR4A}=\cost{NOR4A}+\cost{NOT}.
\end{equation}
Following from Section~\ref{sec:components}, in our implementations, gates with four or less inputs are atomic, while anything larger is composite. However, if gates with more than two inputs are unavailable, the last four equations can be substituted for composite versions. For example, \eqref{eq:C_OR4C} could be used for a composite OR4 gate.
\begin{equation}\label{eq:C_OR4C}
\cost{OR4C}=2\cost{NOR2}+\cost{NAND2}.
\end{equation}

As described in Section~\ref{sec:mlpe}, our PEs use many wide OR gates. For ASIC, these are based on the OR8 unit described in Section~\ref{sec:components:or_gates} and shown in Fig.~\ref{fig:or8u}. The transistor cost of this component is given by \eqref{eq:C_OR8U}:
\begin{equation}\label{eq:C_OR8U}
\cost{OR8U}=4\cost{NOR2}+\cost{NAND4A}.
\end{equation}
To find the cost of larger OR trees constructed from this unit, use \eqref{eq:C_ORn}: 
\begin{equation}\label{eq:C_ORn}
\cost{OR$x$}(x)=\left\{\begin{array}{lr}
        \cost{OR2} + \dfrac{x}{8} \cdot \cost{OR8U} & \text{for } \dfrac{x}{8}\leq 2,\\\\
        \cost{OR4A} + \dfrac{x}{8} \cdot \cost{OR8U} & \text{for } 2<\dfrac{x}{8}\leq 4,\\\\
        \cost{OR8U} + \dfrac{x}{8} \cdot \cost{OR8U} & \text{for } 4<\dfrac{x}{8}\leq 8,\\\\
        \cost{OR$x$}\left(\frac{x}{8}\right) + \dfrac{x}{8} \cdot \cost{OR8U} & \text{for } \dfrac{x}{8}>8.
        \end{array}\right.
\end{equation}
Note that this is the first of these equations that is a (recursive) function; this will be more common moving forward.

\subsubsection{Multiplexer Costs}
The transistor cost of an atomic 1-bit, 2-to-1 mux (or, in our notation, a 2::1 mux), as described in Section~\ref{sec:components:muxes}, is given by \eqref{eq:C_2::1MUXA}, while an alternative composite version is given by \eqref{eq:C_2::1MUXC}.%
\footnote{Here, $\text{M}$ in \eqref{eq:C_2::1MUXA} and subsequent equations serves as an indicator that the equation specifically refers to the complexity of a mux.} 
\begin{equation}\label{eq:C_2::1MUXA}
\cost{2::1M}=8.
\end{equation}
\begin{equation}\label{eq:C_2::1MUXC}
\cost{2::1M}=\cost{NOT}+3\cost{NAND2}.
\end{equation}

Then, \eqref{eq:C_4::1MUXB} can be used if transistor-level granularity is possible, i.e., if using \eqref{eq:C_2::1MUXA}. Otherwise, \eqref{eq:C_4::1MUXA} can be used for the design previously shown in Fig.~\ref{fig:mux4}.
\begin{equation}\label{eq:C_4::1MUXB}
\cost{4::1M}=3\cost{2::1M}.
\end{equation}
\begin{equation}\label{eq:C_4::1MUXA}
\cost{4::1M}=36.
\end{equation}

The transistor cost of an arbitrary single-bit mux with $x$ input bits/channels is given by \eqref{eq:C_1bMUXn}.
\begin{equation}\label{eq:C_1bMUXn}
\cost{$x$::1M}(x)=\left\{\begin{array}{lr}
        \cost{2::1M} + \dfrac{x}{4} \cdot \cost{4::1M} & \text{for } \dfrac{x}{4}\leq 2,\\\\
        \cost{4::1M} + \dfrac{x}{4} \cdot \cost{4::1M} & \text{for } 2<\dfrac{x}{4}\leq 4,\\\\
        \cost{$x$::1M}\left(\frac{x}{4}\right) + \dfrac{x}{4} \cdot \cost{4::1M} & \text{for } \dfrac{x}{4}>4.
        \end{array}\right.
\end{equation}

Finally, the transistor cost of an arbitrary $x::y$ mux, as described in Section~\ref{sec:components:muxes}, is given by \eqref{eq:C_MUXn}.
\begin{equation}\label{eq:C_MUXn}
\cost{$x$::$y$M}(x,\,y)=\frac{x}{y} \cdot \cost{$x$::1M}(y).
\end{equation}

\subsubsection{SLPE Costs}
As detailed in Section~\ref{sec:methods-fpga}, we use mux-based SLPEs. Thus, we can provide a straightforward expression for the complexity of an SLPE of arbitrary size if we know how many single-bit 2:1 muxes it contains. Note that in an SLPE, all muxes have two input channels.%
\footnote{This can be unintuitive; see Fig.~\ref{fig:slpe-8-mux} for a visual example.}
Thus, our other optimizations and decisions for multiplexers do not apply here. With that in mind, \eqref{eq:slpe_mux_count_cost} gives the desired number of multiplexers in an $n:\lfloor\log_2 n\rfloor$ SLPE.
\begin{equation}\label{eq:slpe_mux_count_cost}
M_\text{SLPE-C}(n)=\sum\limits_{i=2}^{\lfloor\log_{2}n\rfloor}\big(i\cdot2^{i-1}\big).
\end{equation}

Then, the transistor cost of the SLPE is given by \eqref{eq:C_SLPE}:
\begin{equation}\label{eq:C_SLPE}
\cost{SLPE}(n)=M_\text{SLPE}(n) \cdot \cost{2::1M}.
\end{equation}

\subsubsection{Recursive PE Costs}
One target for comparison for our ASIC analysis is the recursive design introduced in Section~\ref{sec:recursive_pe}. Based on \cite{bcam}, we developed equations to calculate the cost in transistors of their design, allowing direct and fair comparison to our designs. As recommended by \cite{bcam}, we take the division factor $k$ to be $4$, i.e., we will split the PE into four smaller PEs with each level of recursion.%
\footnote{Note that this parameter $k=4$ is distinct from both the $k$ from \cite{2lmr} which is equivalent to $L_1$ in this work, and the $k=\lfloor\log_2n\rfloor$ used in \cite{mld}. We avoid using $k$ as a variable in this work to avoid ambiguity.}

For brevity, we first define the function $w_R(x)$ to be the mux signal width for a given PE input width $x$ for an arbitrary level of recursion. This is given by 
\begin{equation}\label{eq:w_R}
w_R(x)=\left\lceil\log_2 \frac{x}{k}\right\rceil=\left\lceil\log_2 \frac{x}{4}\right\rceil.
\end{equation}

Then, the transistor cost of the recursive PE is given by \eqref{eq:C_REC}, assuming the base case is an encoder size less than or equal to $k=4$.
\begin{equation}\label{eq:C_REC}
\cost{REC}(n)=\left\{\begin{array}{lr}
        \cost{SLPE}(n) & \text{for } n\leq k,\\\\
        \cost{SLPE}(k)+\\\cost{$x$::1M}\big(4w_R(n),\,w_R(n)\big)+\\k \cdot \cost{REC}\left(\dfrac{n}{k}\right) & \text{for } n>k.
        \end{array}\right.
\end{equation}
Essentially, for each layer until the base encoders, we must calculate the size of the $k:\lfloor\log_2k\rfloor$ (4:2) PE and the $\lceil \log_2 \frac{n}{k}\rceil$ bit wide $k:1$ (4:1) mux, then recurse for the $k=4$ identical sub-encoders.

\subsubsection{Tree PE Costs}
We also compare to the tree PE introduced in Section~\ref{sec:tree_pe}. Similarly to the recursive PE, we developed an appropriate expression for the complexity, given by
\begin{equation}\label{eq:C_TREE}
\cost{TREE}(n)=\left\{\begin{array}{lr}
        \cost{NOT}+\cost{OR2} & \text{for } n\leq 2,\\\\
        \cost{NOT}+\cost{OR2}+\\(\lfloor \log_2n\rfloor-1) \cdot \cost{2::1M}+\\2 \cdot \cost{TREE}\left(\dfrac{n}{2}\right) & \text{for } n>2.
        \end{array}\right.
\end{equation}

\subsubsection{2LPE Costs}
For the 2LPE, the total transistor cost is the sum of that of the OR gates, coarse PE, mux, and fine PE. As described in Section~\ref{sec:2lpe}, there are $L_1$ individual $L_2:1$ OR gates feeding the coarse PE. The mux feeding the fine PE is of size $n::L_2$. Thus, the total cost of the combined circuit is given by
\begin{equation}\label{eq:C_2LPE}
\begin{array}{rl}
\cost{2LPE}(n)=&\!\!\!\!
L_1 \cdot \cost{OR$x$}(L_2)
+ \cost{SLPE}(L_1) \\&+ \cost{$x$::$y$M}(n, L_2) + \cost{SLPE}(L_2).
\end{array}
\end{equation}

\subsubsection{Composed MLPE Costs}
For the composed MLPE, the total transistor cost can be found recursively using the 2LPE as the base case, as shown by \eqref{eq:C_MLPE_O}:
\begin{equation}\label{eq:C_MLPE_O}
\cost{MLPE-O}(n,m)=\left\{\begin{array}{lr}
        \cost{2LPE}(n) & \text{for } m= 2,\\\\
        L_1 \cdot \cost{OR$x$}(L_2)
        +\\
        \cost{MLPE-O}(L_1,\,m-1)+\\
        \cost{$x$::$y$M}(n, L_2)+\\
        \cost{MLPE-O}(L_2,\,m-1)& \text{for } m \geq 3.
        \end{array}\right.
\end{equation}
Effectively, this is the same as \eqref{eq:C_2LPE} except self-recursive. To calculate this value, the original $m$ (e.g., $m=4$ for a 4LPE) should be passed to the function, and the recursive calls will successively decrement this value to calculate the complexity of the component PEs.

\subsubsection{Cascaded MLPE Costs}
Because the recursion of the cascaded variant differs from the composed MLPE, calculating the transistor cost is slightly more complex. First, consider that the number of OR gates in a given level will be the product of all $L_i$ except $L_m$, or, equivalently, $n/L_m$. Each of these OR gates will be $L_m$ bits wide. Refer to Fig.~\ref{fig:4096-3lpe-cascade-diag} for a visual example, and refer to \eqref{eq:mlpe_cascade_L_1} and \eqref{eq:mlpe_cascade_L_i} for definitions of $L_i$. We can then use \eqref{eq:C_ORn} to find the cost of each OR gate, and the cost of the other components can be found by using \eqref{eq:C_MUXn} for the cost of each $n::L_m$ mux and \eqref{eq:C_SLPE} for the level-$m$ sub-encoder. Then, similarly to the cost equation for the composed technique, we use the 2LPE as the base case, and perform recursion down from level $m$. The total cost for a cascaded MLPE is given by \eqref{eq:C_MLPE_A}.
\begin{equation}\label{eq:C_MLPE_A}
\cost{MLPE-A}(n,m)=\left\{\begin{array}{lr}
        \cost{2LPE}(n) & \text{for } m= 2,\\\\
        \frac{n}{L_m} \cdot \cost{OR$x$}(L_m)
        +\\
        \cost{$x$::$y$M}(n, L_m)+\\
        \cost{SLPE}(L_m)+\\
        \cost{MLPE-A}\Big(\frac{n}{L_m},\,m-1\Big)& \text{for } m \geq 3.
        \end{array}\right.
\end{equation}

\subsection{ASIC Delay Analysis}\label{sec:methods-asic2}
This section analyzes the delay, or critical path, of the architectures whose complexity is defined in the previous section. We also provide a separate Desmos calculator for this section here: \href{https://www.desmos.com/calculator/hrn1hfngfd}{www.desmos.com/calculator/hrn1hfngfd}. The critical path is reported as a total number of \q{normalized transistors} defined as follows. Below, we analyze the atomic components of PEs down to the transistor level, and differentiate the delay between NMOS and PMOS transistors using a ratio of 2:1. This means that the delay of one PMOS transistor is considered to be equivalent to that of two NMOS transistors in series. The ratio is configurable in the aforementioned Desmos calculator to allow for application towards arbitrary manufacturing technologies, but is assumed to always be at least 1:1. Notation for equations/functions in this section uses $D$ for total delay (critical path in terms of normalized transistors). Each of these are subscripted as before, corresponding to the specific component, i.e., NOT gate, SLPE, etc.

\subsubsection{Base Logic Gate Delay}
First, we analyze the delay of the fundamental components, as described above. In Table~\ref{tab:gate_delay}, we summarize the critical paths of the basic logic gates used in PEs, first individually as NMOS and PMOS, then as a normalized total at a ratio of 2:1, as described previously.

\begin{table}[ht]
    \centering
    \begin{tabular}{|c|cc:c|}\hline
       \textbf{Gate}  & \textbf{NMOS} & \textbf{PMOS} & \textbf{Norm.} \\\hline
       NOT   & 0 & 1 & 2  \\
       NAND2 & 0 & 1 & 2  \\
       NAND3 & 3 & 0 & 3  \\
       NAND4 & 4 & 0 & 4  \\
       NOR2  & 0 & 2 & 4  \\
       NOR4  & 0 & 4 & 8  \\
       AND2  & 0 & 2 & 4  \\
       OR2   & 0 & 3 & 6  \\
       OR4   & 0 & 5 & 10 \\
       OR8U  & 4 & 2 & 8  \\\hline
    \end{tabular}
    \vspace{1em}
    \caption{Atomic Logic Gate Delays}
    \label{tab:gate_delay}
    \vspace{-1em}
\end{table}
In the following equations, this table will be referred to via subscripted $D$ variables. For example, $\delay{NOT}$ refers to the upper right-hand data cell in the table. Furthermore, note that \q{OR8U} in Table~\ref{tab:gate_delay} refers to the \q{OR8 Unit} structure defined in Section~\ref{sec:components:or_gates}. Building upon this, \eqref{eq:D_ORn} expresses the delay of a tree of OR gates for an arbitrary input width.
\begin{equation}\label{eq:D_ORn}
\delay{OR$x$}(x)=\left\{\begin{array}{lr}
        \delay{OR2} + \delay{OR8U} & \text{for } \dfrac{x}{8}\leq 2,\\\\
        \delay{OR4A} + \delay{OR8U} & \text{for } 2<\dfrac{x}{8}\leq 4,\\\\
        2\cdot \delay{OR8U} & \text{for } 4<\dfrac{x}{8}\leq 8,\\\\
        \delay{OR$x$}\left(\frac{x}{8}\right) + \delay{OR8U} & \text{for } \dfrac{x}{8}>8.
        \end{array}\right.
\end{equation}

\subsubsection{Multiplexer Delay}
For the transistor-level optimized design shown in Fig.~\ref{fig:mux2a} whose complexity is described by \eqref{eq:C_2::1MUXA}, the normalized delay is given by %
\begin{equation}\label{eq:D_2::1MUXA}
\delay{2::1M}=4.
\end{equation}
The critical path here is 2 PMOS transistors. Meanwhile, the delay for the alternative design described by \eqref{eq:C_2::1MUXC} is given by %
\begin{equation}\label{eq:D_2::1MUXC}
\delay{2::1M}=\delay{NOT}+2\delay{NAND2}.
\end{equation}
The critical path for this design is one inverter followed by two NAND2 gates. For the results presented in Section~\ref{sec:results-asic}, we use the former.

Similarly, the delays for the 4::1 mux designs described in \eqref{eq:C_4::1MUXB} and \eqref{eq:C_4::1MUXA} are given by \eqref{eq:D_4::1MUXB} and \eqref{eq:D_4::1MUXA}, respectively. For the results presented in Section~\ref{sec:results-asic}, we use \eqref{eq:D_4::1MUXB}.
\begin{equation}\label{eq:D_4::1MUXB}
\delay{4::1M}=2\delay{2::1M}.
\end{equation}
\begin{equation}\label{eq:D_4::1MUXA}
\delay{4::1M}=\delay{NOT}+\delay{NAND3}+\delay{NAND4}.
\end{equation}

As previously established, we construct muxes of arbitrary width using a fan-in of at most 4. With that in mind, \eqref{eq:D_1bMUXn} gives the delay of a single-bit arbitrary-length mux, corresponding to \eqref{eq:C_1bMUXn}.
\begin{equation}\label{eq:D_1bMUXn}
\delay{$x$::1M}(x)=\left\{\begin{array}{lr}
        \delay{2::1M} + \delay{4::1M} & \text{for } \dfrac{x}{4}\leq 2,\\\\
        2\cdot \delay{4::1M} & \text{for } 2<\dfrac{x}{4}\leq 4,\\\\
        \delay{$x$::1M}\left(\frac{x}{4}\right) + \delay{4::1M} & \text{for } \dfrac{x}{4}>4.
        \end{array}\right.
\end{equation}
Similarly, \eqref{eq:D_MUXn} gives the delay of a multi-bit mux described in our notation, corresponding to \eqref{eq:C_MUXn}. Unfortunately, our notation makes this particular equation slightly counterintuitive, as the delay is affected only by the input length of the single-bit muxes working in parallel to provide a total input length of $n$ bits.
\begin{equation}\label{eq:D_MUXn}
\delay{$x$::$y$M}(x,\,y)=\delay{$x$::1M}(y).
\end{equation}

\subsubsection{SLPE Delay}
Similarly to \eqref{eq:D_MUXn}, although mux-based SLPEs are constructed out of incrementally wider 2:1 muxes, the width is irrelevant to the critical path since all the muxes are in parallel and have the same number of channels (two). Thus, we only need to know the number of 2:1 muxes in the critical path. This turns out to be $n-2$, which follows since that is the number of two-option \q{choices} that must be made for $n$ values. We also verified this empirically up to 64 bits. Finally, the number of arbitrary-width 2:1 muxes is then given by %
\begin{equation}\label{eq:slpe_mux_count_delay}
M_\text{SLPE-D}(n)=n-2.
\end{equation}
Note that \eqref{eq:slpe_mux_count_delay} is \textit{not} equivalent to \eqref{eq:slpe_mux_count_cost} because these 2:1 muxes have varying input widths, whereas the 2:1 muxes counted by \eqref{eq:slpe_mux_count_cost} are all single-bit.

From this, we can determine the delay of the entire SLPE. The critical path will travel through all $n-2$ muxes, but once again, the varying widths of each mux do not affect the critical path. Thus, we simply combine \eqref{eq:slpe_mux_count_delay} with \eqref{eq:D_2::1MUXA}---or \eqref{eq:D_2::1MUXC} if choosing the composite version---to obtain \eqref{eq:D_SLPE}.
\begin{equation}\label{eq:D_SLPE}
\delay{SLPE}(n)=(n-2) \cdot \delay{2::1M}.
\end{equation}

\subsubsection{Recursive PE Delay}
Similar to the analysis in the previous section, using the same parameters---and \eqref{eq:w_R}---we can express the delay of the recursive PE from Section~\ref{sec:recursive_pe} with \eqref{eq:D_REC}.
\begin{equation}\label{eq:D_REC}
\delay{REC}(n)=\left\{\begin{array}{lr}
        \delay{SLPE}(n) & \text{for } n\leq k,\\\\
        \delay{$x$::1M}\big(4w_R(n),\,w_R(n)\big)+\\\delay{REC}\left(\dfrac{n}{k}\right) & \text{for } n>k.
        \end{array}\right.
\end{equation}

\subsubsection{Tree PE Delay}
As before, the tree PE is similar to the recursive PE, so the delay equation is similar. The delay for the tree PE is given by \eqref{eq:D_TREE}.
\begin{equation}\label{eq:D_TREE}
\delay{TREE}(n)=\left\{\begin{array}{lr}
        \delay{OR2} & \text{for } n\leq 2,\\\\
        \delay{2::1M}+\delay{TREE}\left(\dfrac{n}{2}\right) & \text{for } n>2.
        \end{array}\right.
\end{equation}

\subsubsection{2LPE Delay}
For the 2LPE, the critical path is essentially the entire circuit, starting at the OR gates, proceeding through the coarse PE, entering the mux via the select signal (which is the output of the coarse PE), continuing through the fine PE, and finally exiting the circuit. The total critical path in normalized transistors is given by \eqref{eq:D_2LPE}.
\begin{equation}\label{eq:D_2LPE}
\begin{array}{rl}
\delay{2LPE}(n)=&\!\!\!\!\delay{OR$x$}(L_2) + \delay{SLPE}(L_1) \\&+ \delay{$x$::$y$M}(n, L_2) + \delay{SLPE}(L_2).
\end{array}
\end{equation}

\subsubsection{Composed MLPE Delay}
As with complexity, the composed MLPE is nearly identical to the 2LPE in its delay equation. The critical path can be determined recursively using the 2LPE as the base case with \eqref{eq:D_MLPE_O}.
\begin{equation}\label{eq:D_MLPE_O}
\delay{MLPE-O}(n,m)=\left\{\begin{array}{lr}
        \delay{2LPE}(n) & \text{for } m= 2,\\\\
        \delay{OR$x$}(L_2)+\\
        \delay{MLPE-O}(L_1,\,m-1)+\\
        \delay{$x$::$y$M}(n, L_2)+\\
        \delay{MLPE-O}(L_2,\,m-1)& \text{for } m \geq 3.
        \end{array}\right.
\end{equation}

\subsubsection{Cascaded MLPE Delay}
Finally, the equation for the critical path of the cascaded MLPE is given by \eqref{eq:D_MLPE_A}. Unsurprisingly, it is similar to its complexity equation, working recursively with the 2LPE as the base case.
\begin{equation}\label{eq:D_MLPE_A}
\delay{MLPE-A}(n,m)=\left\{\begin{array}{lr}
        \delay{2LPE}(n) & \text{for } m= 2,\\\\
        \cost{OR$x$}(L_m)+\\
        \delay{$x$::$y$M}(n, L_m)+\\
        \delay{SLPE}(L_m)+\\
        \delay{MLPE-A}\Big(\frac{n}{L_m},\,m-1\Big)& \text{for } m \geq 3.
        \end{array}\right.
\end{equation}

\subsection{Replication of Recursive and Tree Architectures}
As mentioned in Section~\ref{sec:bg}, in order to fairly compare the proposed designs against the state-of-the-art designs, we replicated the recursive design proposed by \cite{bcam} (introduced in Section~\ref{sec:recursive_pe}) and the tree design proposed by \cite{tree} (Section~\ref{sec:tree_pe}). The only major change that we made in our implementation was to the recursive design. The Verilog descriptions provided by the authors on their GitHub page were pipelined/registered between stages. Our other designs are currently not pipelined, although this would be possible. Thus, to compare fairly, our reconstruction of the recursive architecture is not pipelined.\footnote{We also used VHDL instead of Verilog.} %
This matches the other tested hardware in that only combinational logic is used.

Aside from that, for FPGA, we implemented both the recursive and tree architectures in VHDL directly according to their original block diagrams, and for ASIC, we derived transistor cost equations similarly. It is also worth noting again that most designs presented are technically recursive, including the tree PE and our MLPEs. However, we attempted to utilize similar descriptions as \cite{bcam} and \cite{tree}. Thus, these names serve primarily to easily distinguish between the different architectures and the works in which they were proposed.

\section{Results and Discussion}\label{sec:results}
In this section, we present the results of our complexity and delay analyses for FPGA and ASIC implementations of the architectures discussed earlier. Then, we analyze the differences in the results between the two implementation technologies and offer recommendations regarding the most suitable architectures for different input precisions, considering the target technology alongside design goals and constraints. The data discussed next is generally summarized by graphs and tables; the full dataset is available at our GitHub repository\footnote{Accessible at: \href{https://github.com/ALUminaries/Multi-Level-Priority-Encoders}{github.com/ALUminaries/Multi-Level-Priority-Encoders}.} for further reference.

\subsection{FPGA Synthesis Results}\label{sec:results-fpga}
First, we report the results of FPGA synthesis, performed as described in Section~\ref{sec:methods-fpga}. The input lengths tested ranged from $n=4$ to $n=2^{18}=262144$ bits, with exceptions noted where applicable. Tested architectures include %
the mux-based SLPE for $n\leq 32768$,\footnote{We elected to halt SLPE testing here since at $n>32768$ would require bypassing the default Vivado loop limit, and larger SLPE data adds little value.} %
the recursive structure from \cite{bcam} introduced in Section~\ref{sec:recursive_pe}, %
the tree structure from \cite{tree} (Section~\ref{sec:tree_pe}), %
the two-level structure for $n\geq16$ (Section~\ref{sec:2lpe}), %
and finally the MLPE paradigm (Section~\ref{sec:mlpe}) for $m=3$ and $m=4$, using the composed and cascaded techniques for $n\geq 512$.%
\footnote{For the 2LPE and MLPE, smaller implementations are impractical or infeasible (e.g., 3LPE for $n\leq256$ would result in $n\leq 4$ base encoders).}

\begin{figure}[h!]
    \centering
    \vspace{-0.75em}
    \includesvg[inkscapelatex=false, width=\columnwidth]{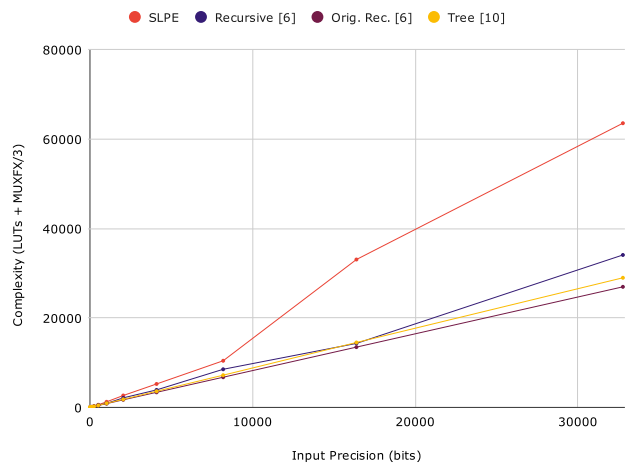} 
    \vspace{-1.75em}
    \caption{FPGA complexity in adjusted synthesis LUTs for existing architectures (Section~\ref{sec:bg}) up to $n=32$ Kib.}
    \label{fig:fpga_cost_existing}
\end{figure}

\begin{figure*}[ht!]
    \centering
    \vspace{-0.75em}
    \includesvg[inkscapelatex=false, width=1.66\columnwidth]{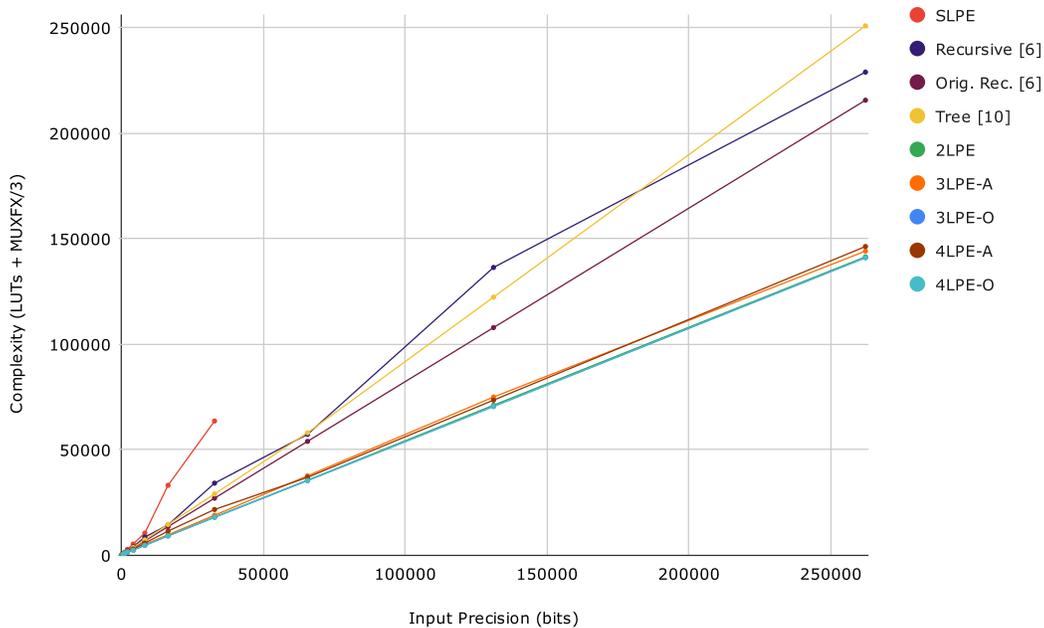} 
    \vspace{-1.25em}
    \caption{FPGA complexity in adjusted synthesis LUTs for all reported architectures.}
    \label{fig:fpga_cost_all}
    \vspace{-0.25em}
\end{figure*}

Fig.~\ref{fig:fpga_cost_existing} shows the complexity (or cost), in adjusted LUTs \eqref{eq:muxfx_lut_conv}, of the existing designs described in Section~\ref{sec:bg}, i.e., the SLPE,  recursive, and tree architectures. Of note, the complexity is reported for the original Verilog description from \cite{bcam}. However, that hardware description is pipelined, using roughly $1.4n$ flip-flops (FFs) in addition to the LUTs used as shown in the figure. This is also why delay is \textit{not} reported for the original description, as it is not directly comparable to the other tested architectures. The reason why the complexity is reported in the first place is to show that our reproduction of the recursive PE is accurate enough to compare our designs against \cite{bcam}. Additionally, the chart shows only up to $n=32768$, to provide a better picture of the behavior at lower input lengths.

Next, Fig.~\ref{fig:fpga_cost_all} shows the complexity in adjusted LUTs \eqref{eq:muxfx_lut_conv} for all reported architectures and input lengths. While Fig.~\ref{fig:fpga_cost_existing} shows a noticeable drop in complexity from SLPE to the recursive and tree PEs, Fig.~\ref{fig:fpga_cost_all} also illustrates that SLPE complexity scales sharply with increasing $n$. From the figures, it is evident that the complexity for each architecture grows roughly linearly, with varying slopes. We also see that the 2LPE architecture provides an appreciable reduction in complexity on top of the recursive and tree PEs. However, contrary to what we might expect, additional levels ($m\geq 2$) do \textit{not} result in any further substantial reductions. 

\begin{figure}[ht]
    \centering
    \vspace{-0.75em}
    \includesvg[inkscapelatex=false, width=\columnwidth]{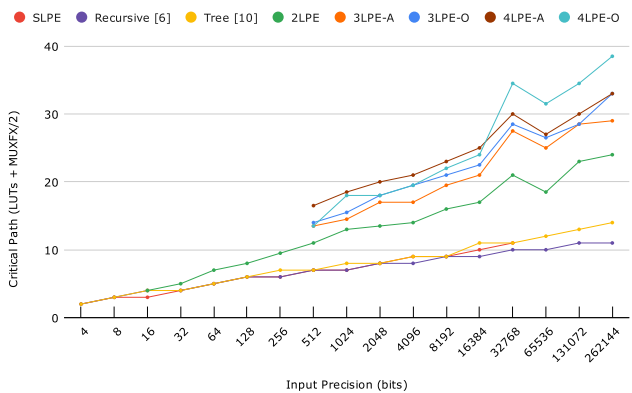} 
    \vspace{-1.5em}
    \caption{FPGA delay (critical path) in adjusted LUT equivalents by architecture. Horizontal axis is logarithmic.}
    \label{fig:fpga_delay_all}
\end{figure}

We now analyze the delay of the FPGA architectures. Like complexity, we report an adjusted value accounting for the MUXFX primitives in the critical path, as described in Section~\ref{sec:methods-fpga} -- in this case, MUXFX count as one-half of a LUT. Fig.~\ref{fig:fpga_delay_all} shows that the SLPEs, recursive PEs, and tree PEs are all roughly the same speed, and overall are the fastest designs tested. The 2LPE has a slightly higher delay, with the gap widening with increasing $n$. 3LPEs and 4LPEs, especially the composed 4LPE, show an additional increase in delay. Interestingly, for all MLPE architectures (including 2LPE), we see a noticeable downtick in delay at $n=65536$. Most likely, this is due to more effective FPGA resource utilization and placement.

\begin{figure}[ht]
    \centering
    \vspace{-0.75em}
    \includesvg[inkscapelatex=false, width=\columnwidth]{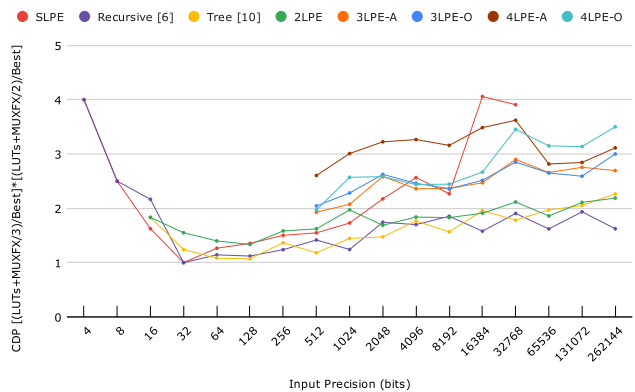} 
    \vspace{-1.5em}
    \caption{FPGA complexity-delay product, relative to best-of-precision, organized by architecture. Horizontal axis is logarithmic. Lower is better.}
    \label{fig:fpga_rcdp}
    \vspace{-0.25em}
\end{figure}

In Fig.~\ref{fig:fpga_rcdp}, we show a metric indicating how strong each architecture is when complexity and delay are considered equally important. Other works commonly use metrics such as the power-delay product or similar. Here, we first find the best (lowest complexity/delay) for each tested input length $n$. Then, independently for both complexity and delay, we divide each data point by that value to obtain a number showing its performance relative to the strongest design for that precision. For instance, the recorded complexity of a 512-bit SLPE was 576 LUTs (adjusted), while a 512-bit 3LPE-A used 372 adjusted LUTs, the lowest for all 512-bit architectures. The relative complexity score for the former is then $576/372\approx 1.55$, while the score of the latter is $372/372=1.0$. This is then multiplied by the corresponding delay score to obtain the value reported in Fig.~\ref{fig:fpga_rcdp}. The lower this value is, the \q{better} the corresponding design is, in terms of complexity and delay. 

Now, we observe the following trends. The efficiency of the reported designs at $n\leq 16$ is relatively poor. In fact, gate-based SLPEs outperform the mux-based ones here, but the focus of this work is on much higher $n$, and other works have extensively covered this input precision range, so we will move on. For $n\geq 32$, we see that the recursive and tree PEs are typically the best, providing a balance between complexity and delay, with the 2LPE closing the gap at $n\geq2048$. Unfortunately, while better in complexity, the MLPE architectures universally have comparatively higher delay due to the additional structure introduced by multiple levels. This highlights a complexity-delay tradeoff, which will be further explored in Section~\ref{sec:results-recs}.

\subsection{ASIC Analysis Results}\label{sec:results-asic}
Next, we report the results of our analyses of complexity and delay for ASIC based on the equations presented in Sections~\ref{sec:methods-asic} and \ref{sec:methods-asic2}. Tested data and architectures were the same as the FPGA data in the previous subsection, except for the following changes: 

\begin{enumerate}[label=\textbf{\arabic*.}]
    \item We did not analyze complexity for the original Verilog code from \cite{bcam}, as that would not make sense in this context.
    \item Range restrictions were technically removed, but certain input lengths for certain designs are effectively ignored, since much like the FPGA results, below a certain $n$, implementing some designs is impractical or infeasible.
    \item We tested composed and cascaded 5LPEs ($m=5$), which are shown with reduced opacity in the graphs, primarily to show why they are not suitable for this range of input precisions.
\end{enumerate} 

\begin{figure}[h!]
    \centering
    \vspace{-0.75em}
    \includesvg[inkscapelatex=false, width=\columnwidth]{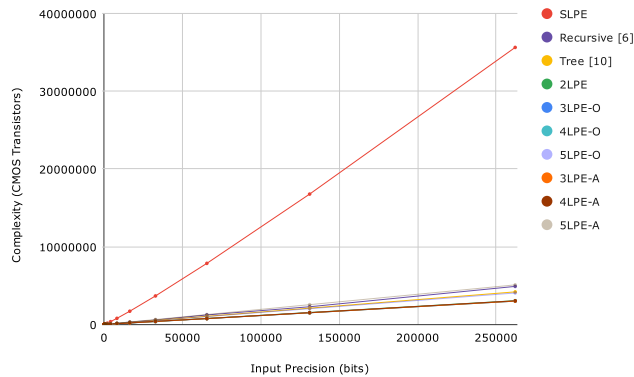} 
    \vspace{-1.5em}
    \caption{ASIC complexity, in static CMOS transistors, for all reported architectures.}
    \label{fig:asic_cost_all}
\end{figure}

\begin{figure}[ht]
    \centering
    \vspace{-0.75em}
    \includesvg[inkscapelatex=false, width=\columnwidth]{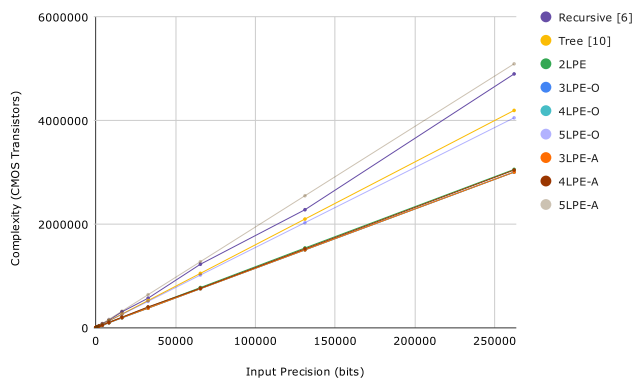} 
    \vspace{-1.5em}
    \caption{ASIC complexity, in static CMOS transistors, for all reported architectures except SLPE.}
    \label{fig:asic_cost_no_slpe}
    \vspace{-0.25em}
\end{figure}

Fig.~\ref{fig:asic_cost_all} shows the complexity (or cost) in static CMOS transistors for all reported architectures and input lengths. It is immediately obvious that the SLPE is by far the most complex design. Thus, in Fig.~\ref{fig:asic_cost_no_slpe}, we omit the SLPE data, providing a better picture of the complexity growth of the architectures. From this, we see a similar picture to the FPGA data. The recursive structure \cite{bcam} has higher complexity than the tree structure \cite{tree} which in turn has higher complexity than the MLPEs for $m<5$. Once again, like the FPGA data, the complexities of those MLPEs exhibit little variance. In fact, the percentage difference of the 3LPEs and 4LPEs from the 2LPE does not exceed 10\% in all reasonable cases.%
\footnote{As noted previously, at lower input lengths, some of the MLPEs are suboptimal. In this particular case, we are looking only at 3LPEs for $n\geq 128$, composed 4LPEs for $n\geq 512$, and cascaded 4LPEs for $n\geq4096$.} 
However, we notice that, excluding the SLPE, the complexity of the cascaded 5LPE is generally the highest, slightly above the recursive PE. The composed 5LPE maintains a similar distance below the tree PE, i.e., the composed 5LPE is slightly less complex. Even so, all of the architectures presented provide a substantial improvement over the SLPE, which is once again clear from Fig.~\ref{fig:asic_cost_all}.

\begin{figure}[ht]
    \centering
    \vspace{-0.75em}
    \includesvg[inkscapelatex=false, width=\columnwidth]{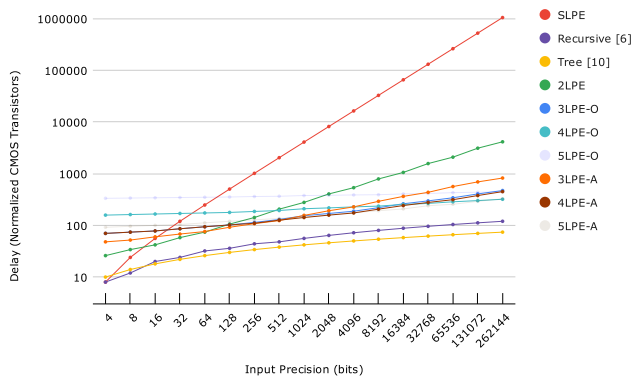} 
    \vspace{-0.75em}
    \caption{ASIC delay (critical path), in static CMOS transistors, for all reported architectures. Both axes are logarithmic.}
    \label{fig:asic_delay_all}
    \vspace{-0.25em}
\end{figure}

\begin{figure}[ht]
    \centering
    \includesvg[inkscapelatex=false, width=\columnwidth]{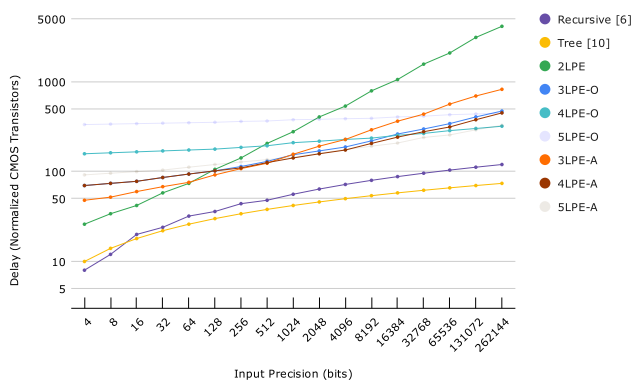} 
    \vspace{-1.5em}
    \caption{ASIC delay (critical path), in static CMOS transistors, for all reported designs except SLPE. Both axes are logarithmic.}
    \label{fig:asic_delay_no_slpe}
    \vspace{-0.25em}
\end{figure}

We now move on to analyze the delay of the ASIC designs. Fig.~\ref{fig:asic_delay_all} and Fig.~\ref{fig:asic_delay_no_slpe} are plots with and without the SLPE data, respectively, similar to the complexity graphs. In both plots, the horizontal and vertical axes are both logarithmic, to help visually separate the data. Surprisingly, and contrary to what we would expect based on the FPGA data, we find that the SLPE has the fastest growth in delay, and the highest for $n\geq 128$. Although, considering the complexity of the SLPE, it is perhaps more curious that its FPGA counterpart was as fast as the recursive and tree architectures.

In Fig.~\ref{fig:asic_delay_no_slpe}, we can see a fairly similar picture to the FPGA delay shown in Fig.~\ref{fig:fpga_delay_all} for the recursive, tree, and 2LPE designs. However, with $m>2$, we actually see a substantial \textit{downtick} in delay, once again contrary to what we would expect based on the FPGA data. In the next subsection, Section~\ref{sec:results-dim-ret}, we will discuss the effects of increasing $m$ on our design factors, and in Section~\ref{sec:results-recs}, we will summarize our findings from the preceding three subsections to provide our usage recommendations.

\subsection{Diminishing Returns}\label{sec:results-dim-ret}
Based on the initial reduction in complexity from the 2LPE design over the traditional SLPE, one might think that further increasing $m$ might yield a similar drop in complexity. As we have seen in this section, that is unfortunately not the case. Compared to the 2LPE, the 3LPEs and 4LPEs achieve only a slight reduction in complexity, for both FPGA and ASIC. In almost all cases, the variance between these architectures is less than 10\%, and it is nearly always a reduction in complexity. However, what we see from the 5LPEs is a clear indication of diminishing returns. This is most evident in Fig.~\ref{fig:asic_cost_no_slpe}, where the data for MLPEs for $2\leq m \leq 4$ are tightly grouped, but the data for $m=5$ grows much faster, as identified in the previous subsection. 

In terms of delay, we observe some interesting patterns. For FPGA, it is clear that increasing $m$ is directly proportional to increased delay, shown best by Fig.~\ref{fig:fpga_delay_all}. However, for ASIC, we see almost the opposite. While less obvious in Fig.~\ref{fig:asic_delay_all} and Fig.~\ref{fig:asic_delay_no_slpe} due to the ranges of the data and the two logarithmic axes, the SLPE's delay grows linearly, while all other architectures have logarithmic delay growth. Thus, aside from the SLPE, the delay of the 2LPE grows fastest. Looking from $n\geq 512$, except for the composed 5LPE, all other architectures have lower delay than the 2LPE. For the cascaded technique, we observe a substantial decrease in delay by switching from $m=3$ to $m=4$ or $m=5$ for $n\geq1024$. Then, for $n\geq 8192$, the cascaded 5LPE is faster than the cascaded 4LPE. The story for the composed MLPEs is more interesting. The composed 3LPE almost exactly matches the cascaded 3LPE, being just barely slower for $n\geq512$. However, further increasing $m$ for the composed technique comes with substantial increases in delay. The composed 4LPE only reaches parity with the cascaded 5LPE at $n=2^{18}$. The composed 5LPE behaves similarly, reaching parity with the cascaded 4LPE and composed 3LPE at the same $n$.

The other major adjustable parameter is $k$, where $k=4$ for the recursive PE (Section~\ref{sec:recursive_pe}) and $k=2$ for the tree PE (Section~\ref{sec:tree_pe}). While the complexity and delay are not substantially different for FPGA, the complexity and delay of the recursive PE are slightly higher for ASIC. While testing with higher values of $k$ would be necessary to be certain, this suggests that higher values for $k$ may result in higher complexity and delay. With that said, it does not seem possible to further reduce complexity or delay by adjusting $k$, since $k=1$ would effectively be an SLPE. However, uneven divisions, different from any of the architectures described here, could result in improvements. 

This idea is supported by the following observation: for MLPEs, the majority of the complexity no longer comes from the atomic encoders, but from the connecting hardware, i.e., the muxes and OR gates.%
\footnote{Relatedly, muxes of the same total input width, but different output width (e.g., same $x$, different $y$ for an $x\!::\!y$ mux) do not seem to differ substantially in terms of complexity. For the mux sizes used in our PEs, on FPGA, the difference in LUTs was less than 3\%, and for ASIC, the difference in transistors was less than 13\%.}
This may not be the case for the recursive and tree designs, as each level of recursion essentially adds a small mux and a small PE. The large amount of small PEs at the base level would likely contribute the most complexity. 

A final note about complexity is that the balance of the subcomponents of MLPEs substantially differs from their FPGA counterparts. Generally, for MLPEs on FPGA, the muxes account for about 53\% to 63\% of the complexity, while the OR gates account for roughly 35\% to 45\%, and the remaining amount (often less than 2\%) is incurred by the actual atomic PEs. However, on ASIC, the muxes typically account for nearly 65\% to 70\% of the complexity, with the OR gates consuming only 25\% to 30\%, and the PEs normally costing less than 5\%. The added complexity of muxes on ASIC is most likely explained by the ease of implementing wide muxes on FPGA, with the ability to implement up to a 16::1 mux on a single logic slice, as discussed in Section~\ref{sec:components:fpga}. Now, our findings thus far will inform our recommendations in the next subsection.

\subsection{Recommendations}\label{sec:results-recs}
The final major contribution of this paper is a suite of recommendations guiding hardware designers in selecting the appropriate PE architecture based on:
\begin{enumerate}[label=\textbf{\arabic*.}]
    \item the design factors they wish to optimize;
    \item the implementation technology (FPGA or ASIC);
    \item the input precision ($n$).
\end{enumerate}
We will first provide recommendations where low complexity is the primary focus, then the same for delay, and finally for a balance of the two.

\subsubsection{To Optimize Complexity}
First, as discussed in Section~\ref{sec:seq_pe}, sequential (or serial) PEs are obviously superior in complexity to any of the architectures tested in this work, at the cost of increased delay. With that said, this category is restricted in scope to combinational PEs only. Table~\ref{tab:lowest_cost} reports the design with the lowest complexity out of all tested architectures for each input precision, separated between FPGA and ASIC. As a reminder, the `-A' and `-O' suffixes refer to the cascaded and composed MLPE designs, respectively.

\begin{table}[h]
\caption{Recommendations for Lowest Complexity}
\label{tab:lowest_cost}
\vspace{-1.25em}
\def\arraystretch{1.15} 
\begin{center}
\begin{tabular}{ |c|c|c| }
\hline
\textbf{Size (bits)} & \textbf{FPGA} & \textbf{ASIC} \\
\hline
64:6        & 2LPE      & 2LPE      \\
128:7       & 2LPE      & 3LPE-O    \\
256:8       & 2LPE      & 3LPE-O    \\
512:9       & 3LPE-A    & 3LPE-O    \\
1024:10     & 4LPE-O    & 3LPE-O    \\
2048:11     & 2LPE      & 3LPE-O    \\
4096:12     & 4LPE-O    & 3LPE-O    \\
8192:13     & 4LPE-O    & 4LPE-O    \\
16384:14    & 4LPE-O    & 4LPE-O    \\
32768:15    & 3LPE-O    & 3LPE-A    \\
65536:16    & 3LPE-O    & 3LPE-A    \\
131072:17   & 4LPE-O    & 4LPE-O    \\
262144:18   & 4LPE-O    & 4LPE-O    \\
\hline
\end{tabular}
\end{center}
\end{table}

In Table~\ref{tab:lowest_cost}, we report the best architecture for 64:6 PEs and above, since our focus is on higher input precisions, which is where we see the largest gains from the tested designs. As we can see, 
the composed MLPEs are most commonly the least complex option. In some cases, the cascaded 3LPE has a slight advantage; however, as discussed in the previous subsection, the variance between the architectures is incredibly low. Thus, what this data essentially shows is that small improvements in complexity can be obtained at higher input lengths by increasing $m$. Finally, while a hardware designer might follow Table~\ref{tab:lowest_cost} explicitly, it is more likely that our other recommendations will be more helpful.

\subsubsection{To Optimize Delay}
In general, a hardware designer might expect that prioritizing delay would come at a potentially substantial cost in terms of complexity. As we have established, that trade-off is inherent to combinational priority encoders, although the tested architectures all provide substantial improvements compared to the SLPE. In Table~\ref{tab:lowest_delay}, we report the architectures with the lowest delay of all tested designs for each input precision according to implementation technology.

\begin{table}[h]
\caption{Recommendations for Lowest Delay}
\label{tab:lowest_delay}
\vspace{-1.25em}
\def\arraystretch{1.15} 
\begin{center}
\begin{tabular}{ |c|c|c| }
\hline
\textbf{Size (bits)} & \textbf{FPGA} & \textbf{ASIC} \\
\hline
64:6        & SLPE/Rec./Tree & Tree \\
128:7       & SLPE/Rec./Tree & Tree \\
256:8       & SLPE/Recursive & Tree \\
512:9       & SLPE/Rec./Tree & Tree \\
1024:10     & SLPE/Recursive & Tree \\
2048:11     & SLPE/Rec./Tree & Tree \\
4096:12     & Recursive      & Tree \\
8192:13     & SLPE/Rec./Tree & Tree \\
16384:14    & Recursive      & Tree \\
32768:15    & Recursive      & Tree \\
65536:16    & Recursive      & Tree \\
131072:17   & Recursive      & Tree \\
262144:18   & Recursive      & Tree \\
\hline
\end{tabular}
\end{center}
\end{table}

Table~\ref{tab:lowest_delay} confirms that, as mentioned in Section~\ref{sec:recursive_pe}, the recursive PE from \cite{bcam} is highly suitable for implementation on FPGA. For every input length, the recursive PE always either has the lowest delay (by one or more LUTs) or is tied with one or both of the SLPE and tree designs. Conversely, on ASIC, the SLPE has a high delay, as we found previously. The recursive PE still does well here, but is consistently beaten by the tree PE from \cite{tree}. At higher precisions, the delay of the tree PE is nearly half that of the recursive PE. Thus, for optimal delay, we recommend the \textbf{recursive design} \cite{bcam} for FPGA, and the \textbf{tree design} \cite{tree} for ASIC.

\subsubsection{To Balance Complexity and Delay}
We will now provide recommendations based on a more balanced and holistic approach to the complexity-delay tradeoff that we have already identified. For FPGA, we inform our approach with the relative complexity-delay product (RCDP) shown in Fig.~\ref{fig:fpga_rcdp}. For ASIC, the numerical data is clearer, so we provide more conclusive remarks. Our recommendations are shown in Table~\ref{tab:balanced} below and explained thereafter.

\begin{table}[h]
\caption{Recommendations for Balanced Complexity \& Delay}
\label{tab:balanced}
\vspace{-1.25em}
\def\arraystretch{1.15} 
\begin{center}
\begin{tabular}{ |c|c|c| }
\hline
\textbf{Size (bits)} & \textbf{FPGA} & \textbf{ASIC} \\
\hline
64:6        & Tree      & Tree          \\
128:7       & Tree      & Tree/3LPE-O   \\
256:8       & Recursive & Tree/3LPE-O   \\
512:9       & Tree      & 3LPE-O        \\
1024:10     & Recursive & 3LPE-O        \\
2048:11     & Tree      & 3LPE-O        \\
4096:12     & Recursive & 3LPE-O        \\
8192:13     & Tree      & 3LPE-O/4LPE-O \\
16384:14    & Recursive & 4LPE-O \\
32768:15    & Tree      & 4LPE-O \\
65536:16    & Recursive & 4LPE-O \\
131072:17   & Recursive & 4LPE-O \\
262144:18   & Recursive & 4LPE-O \\
\hline
\end{tabular}
\end{center}
\end{table}

Generally speaking, using the \textbf{tree design} will be, if not the most optimal choice, certainly a reasonable choice, for any $n$ on either FPGA or ASIC. In Table~\ref{tab:balanced}, for FPGA, we list the architecture with the strictly lowest RCDP. As we can see, the tree PE is commonly the victor there; if it is not, it is usually the second best, although for some values of $n$, the 2LPE is the runner-up. For FPGA, if complexity is \textit{more} important than delay, then the \textbf{2LPE is also a good choice}, as it provides the complexity improvement of the MLPE-type designs without sacrificing too much in terms of delay. Unfortunately, it seems that FPGA routing is not kind to the critical path of the MLPEs, as all $m>2$ designs have a much higher RCDP. Conversely, direct ASIC implementation is much more amenable to those designs. As discussed previously, the MLPEs for $m>2$ see a substantial \textit{decrease} in delay on ASIC. Therefore, we are comfortable recommending the \textbf{composed 3LPE or 4LPE} for $128 \leq n \leq 8192$ or $8192\leq n \leq 262\,144$, respectively. In Table~\ref{tab:balanced} above, where two designs are listed for one input length, they are relatively close in both complexity and delay. In particular, the left has slightly less delay, while the right has slightly less complexity. More generally, on ASIC designs where $n\geq 512$, we see an improvement in complexity of about 21\% to 28\% for the 3/4LPE-O (whichever is recommended for a given $n$) compared to the tree PE. The cost of this improvement is that the corresponding MLPE architecture has between 3.4 and 4.4$\times$ the delay of the tree PE. However, compared to the calculated critical path of the SLPE on ASIC, this is an increase of at most five percent at precisions for $n<4096$, and less than one percent for $n\geq4096$. We believe that this tradeoff is worthwhile, but the hardware designer will need to judge for themselves based on the requirements of their particular project.

\section{Conclusion}\label{sec:conclusion}
This work provides a comprehensive overview of several priority encoder designs, both existing and novel -- namely, the traditional single-level priority encoder, a recursive design, a tree-based design, a two-level priority encoder, and two new multi-level priority encoder structures. After establishing the base components of the latter, we summarized its hardware description for FPGA synthesis, then provided several equations and tools to calculate the complexity and delay of each design for ASIC. Data analysis revealed notable trends in both complexity and delay for both FPGA and ASIC. Typically, the recursive and tree PEs had the least delay, and the MLPEs were the least complex. The data revealed a clear complexity-delay tradeoff between the various architectures, and it also informed our suite of recommendations, serving as a framework to assist hardware designers in selecting the most appropriate architecture.

Future work might explore how pipelining could improve the performance of multi-level priority encoders, especially for high-throughput applications. Another potential research direction may involve evaluating different methods of dividing or balancing the components to further reduce complexity without as much of an impact on delay. For ASIC specifically, one avenue to reduce complexity or delay might come from integrating dynamic CMOS logic or performing lower-level circuit optimizations. Finally, future work could build upon the foundation provided by this work by comparing combinational priority encoders more generally to sequential priority encoders, to formally quantify the difference in both complexity and delay, and better inform hardware designers.

\bibliographystyle{./IEEEtran} 
\bibliography{./IEEEabrv,./references}

\end{document}